\documentclass[reprint,amsmath,amssymb,aps,prb,
               superscriptaddress,a4paper,showkeys]{revtex4-1}
\pdfoutput=1

\usepackage[utf8x]{inputenc}
\usepackage{graphicx}
\usepackage{amsmath}
\usepackage{dcolumn}
\usepackage{wrapfig}
\usepackage{upgreek}
\usepackage{float}
\usepackage{newfloat}
\usepackage{hyperref}
\hypersetup{
  final,
  pdfborder={0 0 0},
  pdftitle={Influence of Crystalline Nanoprecipitates on Shear-Band
    Propagation in Cu-Zr-Based Metallic Glasses},
  pdfauthor={T. Brink, M. Peterlechner, H. Rösner, K. Albe, G. Wilde},
  colorlinks=true,
  urlcolor=blue,
  linkcolor=blue,
  citecolor=blue
}

\makeatletter
\def\ignorecitefornumbering#1{%
     \begingroup
         \@fileswfalse
         #1
    \endgroup
}
\makeatother

\DeclareFloatingEnvironment[name=VIDEO]{myvideo}

\begin{document}
\frenchspacing

\title{Influence of Crystalline Nanoprecipitates on Shear-Band
  Propagation in Cu-Zr-Based Metallic Glasses}

\author{Tobias Brink}
\email{brink@mm.tu-darmstadt.de}
\affiliation{Fachgebiet Materialmodellierung, Institut f{\"u}r
  Materialwissenschaft, Technische Universit\"at Darmstadt,
  Jovanka-Bontschits-Straße~2, D-64287 Darmstadt, Germany}

\author{Martin Peterlechner}
\affiliation{Institut f\"ur Materialphysik, Westf\"alische
  Wilhelms-Universit\"at M\"unster, Wilhelm-Klemm-Stra\ss{}e 10,
  D-48149 M\"unster, Germany}

\author{Harald R\"osner}
\affiliation{Institut f\"ur Materialphysik, Westf\"alische
  Wilhelms-Universit\"at M\"unster, Wilhelm-Klemm-Stra\ss{}e 10,
  D-48149 M\"unster, Germany}

\author{Karsten Albe}
\affiliation{Fachgebiet Materialmodellierung, Institut f{\"u}r
  Materialwissenschaft, Technische Universit\"at Darmstadt,
  Jovanka-Bontschits-Straße~2, D-64287 Darmstadt, Germany}

\author{Gerhard Wilde}
\affiliation{Institut f\"ur Materialphysik, Westf\"alische
  Wilhelms-Universit\"at M\"unster, Wilhelm-Klemm-Stra\ss{}e 10,
  D-48149 M\"unster, Germany}

\date{May 6, 2016}

\begin{abstract}
  \begin{center}
    \makeatletter
    \rule{\frontmatter@abstractwidth}{0.4pt}
    \makeatother
  \end{center}
  \vspace{-0.5\baselineskip}
  The interaction of shear bands with crystalline nanoprecipitates
  in Cu-Zr-based metallic glasses is investigated by a combination
  of high-resolution TEM imaging and molecular-dynamics computer
  simulations.  Our results reveal different interaction mechanisms:
  Shear bands can dissolve precipitates, can wrap around crystalline
  obstacles, or can be blocked depending on the size and density of the
  precipitates. If the crystalline phase has a low yield strength,
  we also observe slip transfer through the precipitate.  Based on
  the computational results and experimental findings, a qualitative
  mechanism map is proposed that categorizes the various processes
  as a function of the critical stress for dislocation nucleation,
  precipitate size, and distance.
  \vspace{-0.75\baselineskip}
  \begin{center}
    \makeatletter
    \rule{\frontmatter@abstractwidth}{0.4pt}
    \makeatother
  \end{center}
  \noindent
  \footnotesize
  Published in:\\
  \href{http://dx.doi.org/10.1103/PhysRevApplied.5.054005}
  {T.~Brink \textit{et al.},
    Phys.\ Rev.\ Applied \textbf{5}, 054005 (2016)}
  \hfill
  DOI: \href{http://dx.doi.org/10.1103/PhysRevApplied.5.054005}
  {10.1103/PhysRevApplied.5.054005}\\[0.5\baselineskip]%
  This article is available under the terms of the
  \href{http://creativecommons.org/licenses/by/3.0/}{Creative
    Commons Attribution 3.0 License}. Further distribution of this
  work must maintain attribution to the authors and the published
  article's title, journal citation, and DOI.
\end{abstract}


\maketitle

\section{Introduction}
\label{sec:intro}

Metallic glasses (MGs) have advantageous mechanical properties, such as
a high yield strength and a large elastic limit, but suffer from brittle
failure, especially under tension, at temperatures significantly below
the glass transition.\cite{Inoue2000,Ashby2006} Under compression,
improved ductility is found for composites of MGs and crystalline
secondary phases, namely, for
Cu-Zr-based,\cite{Fan1997,Lee2006,Hajlaoui2007, Fornell2010,Li2011}
Cu-Ti-based,\cite{Calin2003} and Zr-Ti-based
MGs.\cite{Hays2000,Hays2001,Hofmann2008} Under tension, a small
ductility with 1\% to 2\% strain is observed for Cu-Zr composites
containing nanocrystals.\cite{Pauly2010,Barekar2010,Pauly2010a} With
a higher volume fraction of the crystalline phase, not only compressive
but also significant tensile ductility is reported for
\mbox{Zr-Ti}-based,\cite{Hofmann2008,Qiao2013} Ti-based,\cite{Kim2011}
and Cu-Zr-based MGs.\cite{Liu2012,Wu2014}

For dendritic precipitates, there is a correlation between the
location of dendrites and the occurrence of shear-band
patterns.\cite{Hays2000,Hays2001} The improved ductility is generally
ascribed to the increased number of shear bands and their limited
length given by the constraints of the crystalline
phase.\cite{Hofmann2008} Thus, a high volume fraction of ductile
crystalline phase improves the ductility in compression and
microindentation tests, while a brittle secondary phase does
not.\cite{Fu2007} This is confirmed by Song \textit{et al.}, who
suggest a crystalline volume fraction between $40\%$ and $80\%$ in
Cu-Zr-based MGs for obtaining good mechanical
properties,\cite{Song2013} which is consistent with the fact that
tensile ductility is observed only in glasses with high volume
fractions of ductile crystalline
phases.\cite{Hofmann2008,Kim2011,Liu2012,Wu2014}

While the enhancement of macroscopic ductility of MGs with high volume
fractions of a ductile crystalline phase can be explained by simple
composite models, the influence of nanoprecipitates on the mechanical
properties of MGs containing a much lower volume fraction of
crystalline matter is still not clear.
Similar to the case of dendritic precipitates, shear-band patterns
were also observed in glasses containing small spherical crystallites
with sizes around $2\,\mathrm{nm}$.\cite{Hajlaoui2007}
These nanoprecipitates can grow during deformation in certain metallic
glasses.\cite{Lee2006, Cao2007, Pauly2010, Barekar2010, Pauly2010a,
  Fornell2010} An increased growth rate of nanocrystallites in shear
bands is observed,\cite{Chen2006} which has been related to enhanced
atomic mobility inside shear bands.\cite{Wilde2011,Bokeloh2011}
Deformation-grown nanocrystallites are observed to contain
twins,\cite{Cao2007,Pauly2010,Pauly2010a} which occur only in larger
crystallites, e.g., with a size greater than $20\,\mathrm{nm}$ in a
Cu-Zr-Al MG.\cite{Pauly2010a} These deformation-grown precipitates are
the possible reason for strain hardening during nanoindentation,
\cite{Fornell2010} as well as increased plastic strain during
compression.\cite{Lee2006} It is proposed that the participation of
the crystallites in the plastic deformation is the reason for the
enhanced ductility: Wu \textit{et al.}\ demonstrate that reducing the
stacking-fault energy of B2 CuZr by alloying leads to increased
twinning and higher ductility under tension.\cite{Wu2012} Pauly
\textit{et al.}\ propose that a martensitic transformation from the
B2 phase to the B$19'$ phase with a subsequent volume change is
responsible for toughening in Cu-Zr-based metallic
glasses.\cite{Pauly2010} This interpretation, however, is not
generally accepted.  Corteen \textit{et al.}, in contrast, note that
the volume change of the martensitic transformation is very small and
cannot contribute significantly to toughening.\cite{Corteen2011} They
instead suggest that precipitates increase plasticity by favoring
the nucleation of new shear bands over the growth of critical shear
bands.
Indeed, recent simulation and experimental results provide evidence
for the fact that crystal--glass interfaces serve as nucleation sites
for shear bands and are therefore responsible for the simultaneous
nucleation of multiple shear
bands.\cite{Albe2013,Zaheri2014,Wang2014b} In tensile tests and
corresponding molecular dynamics (MD) simulations of
na\-no\-lam\-inates of copper nanocrystals separated by thin Cu-Zr
glass layers, the crystal--glass interface acts as a source or sink for
dislocations.  Shear transformation zones (STZs) are activated by
interactions with dislocations.\cite{Wang2007,Arman2011,Brandl2013}

\begin{figure}[b]
  \centering
  \includegraphics{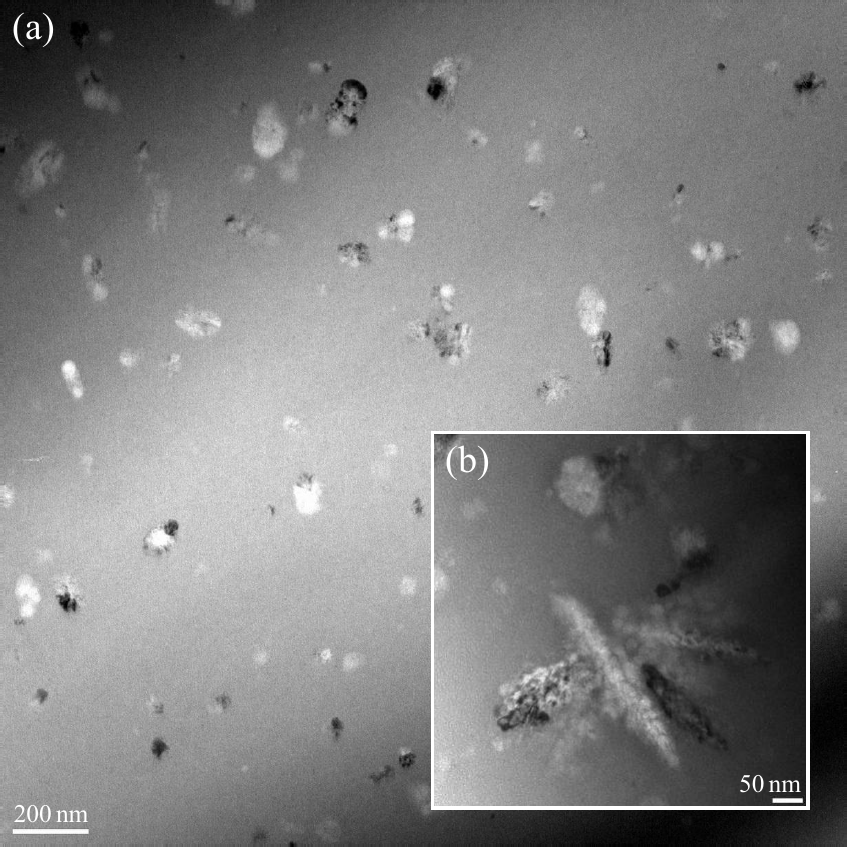}
  \caption{An undeformed Zr$_{53.8}$Cu$_{31.6}$Ag$_{7.0}$Al$_{7.6}$
    sample after annealing in a TEM bright-field image. An overview of
    the sample is shown in (a), with clearly visible crystalline
    precipitates. The inset (b) shows a magnified view, indicating the
    transition from globular to dendritic morphologies during
    precipitate growth. The different brightness of parts of one
    precipitate in (b) is most likely due to the different geometrical
    orientations of different parts of the same precipitate with
    respect to the incident electron beam. The strong black-white
    contrast change observed for some nanocrystals in (a) is most
    likely due to twinning.}
  \label{fig:tem-overview}
\end{figure}
Computational studies on the interaction of crystalline precipitates
with shear bands provide further insights into nanoscale mechanisms.
Lund and Schuh conduct quasi-2D molecular-statics simulations of a
binary Len\-nard-Jones system with a nanocrystal
inclusion.\cite{Lund2007} They identify three mechanisms of
deformation, depending on the ratio of shear-band thickness to crystal
size. For small crystals, the deformation is accommodated either in
the interface (for example, by rotation) or by dissolution of the
crystal. For wide shear bands and intermediate crystal sizes,
dislocations in the crystal nucleate at the interface. Finally, for
crystals larger than the shear band, they observe homogeneous
dislocation nucleation due to stress building in the nanocrystal
center. However, it is somewhat unclear how the observed homogeneous
dislocation nucleation depends on the artificially induced shear band
and the resulting stress state in the system.  Shi and Falk conduct
molecular-dynamics simulations on a monoatomic amorphous model system
with a high fraction of bcc nanocrystallites.\cite{Shi2008} They find
that deformation is induced at the interfaces and that shear bands
bend around crystallites away from a direction of maximum resolved
shear stress. They also observe blocking of shear bands by
crystallites. Because of the high fraction of crystalline phase, the
system more closely resembles a nanocrystalline structure. The
observation of the initiation of plastic deformation at interfaces still
matches the simulations by Albe \textit{et al.}\ \cite{Albe2013} and
underlines the importance of the crystal--glass interface in these
composite systems.

While it has been shown that interfaces promote shear-band nucleation
and that precipitates can act as obstacles or can deform together with
the matrix, there is no comprehensive study that investigates the
influence of the size and number density of the precipitates. Furthermore,
some mechanisms governing the interaction between a propagating shear
band and a preexisting precipitate have been observed but not
investigated and discussed in detail. Therefore the goal of this study
is to investigate the interaction of a shear band with preexisting
precipitates in Cu-Zr-based MGs.  In the experimental part of the
study, we anneal Zr-Cu-Ag-Al melt-spun ribbons to induce the
formation of nanocrystalline precipitates.  We present transmission
electron microscopy (TEM) images of the samples before and after
deformation by cold rolling and identify the effects of crystalline
precipitates on the shear-band propagation.  Using molecular-dynamics
computer simulations, we model composite systems with a metallic-glass
matrix and crystalline precipitates.  We control the initiation of a
shear band using a stress concentrator and put precipitates in its
propagation path.  The focus of the simulations is on the size effects
of ``hard'' precipitates that do not partake in the plastic
deformation. Additionally, we study the shear-band interaction with
``soft,'' plastically deformable precipitates.  Finally, we derive a
deformation map from the combined observations of simulations and
experiments that classifies the observed mechanisms.

\section{Experiment}
\label{sec:experiment}

\subsection{Experimental setup}

We prepare metallic-glass samples of nominal composition
Zr$_{53.8}$Cu$_{31.6}$Ag$_{7.0}$Al$_{7.6}$ from pure components (Cu:
99.999\%, Zr: 99.998\%, Ag: 99.999\%, Al: 99.999\%; all in at.\%) by
prealloying using arc melting. After repeated arc melting with
intermittent turning of the specimen to enhance homogenization, the
entire ingots are inserted into quartz-glass crucibles for melt
spinning. The weight loss during alloying is minimal and subsequent
composition analyses by energy-dispersive x-ray diffraction (EDX)
confirm that the composition of the material is equal to the nominal
composition within the accuracy of the measurement. For melt spinning,
the ingots are inductively melted under an Ar atmosphere and the melt
is ejected onto a rotating Cu wheel (tangential wheel velocity:
$30\,\mathrm{m/s}$), resulting in completely amorphous thin ribbon
samples of approximately $80$-$\mathrm{\upmu{}m}$ thickness. X-ray
diffraction on the as-quenched ribbon samples does not indicate the
presence of any crystalline fraction exceeding the sensitivity
threshold of this method. Parts of the ribbon samples are cut to
perform differential scanning calorimetry (DSC) measurements in a
Perkin Elmer Diamond DSC device. Both isochronal and isothermal
measurements under a purified Ar gas flow are conducted and in
conjunction with microstructure analyses the time and temperature
dependence of the evolving crystalline fraction is determined. On the
basis of these results, the samples for deformation processing are
annealed in the DSC device at $410\,^\circ\mathrm{C}$ for
$3\,\mathrm{h}$. This thermal treatment results in the formation of
nanocrystalline precipitates with an average diameter of about
$70\,\mathrm{nm}$ and a number density on the order of
$10^{20}\,\mathrm{m^{-3}}$. The resulting microstructure is shown in
Fig.~\ref{fig:tem-overview}.  Given that the width of shear bands is on the
order of $10\,\mathrm{nm}$,\cite{Roesner2014} there is no straight
path for propagating shear bands to avoid the interaction with
nanoprecipitates in these samples.

\begin{figure}
  \centering
  \includegraphics[width=8.6cm]{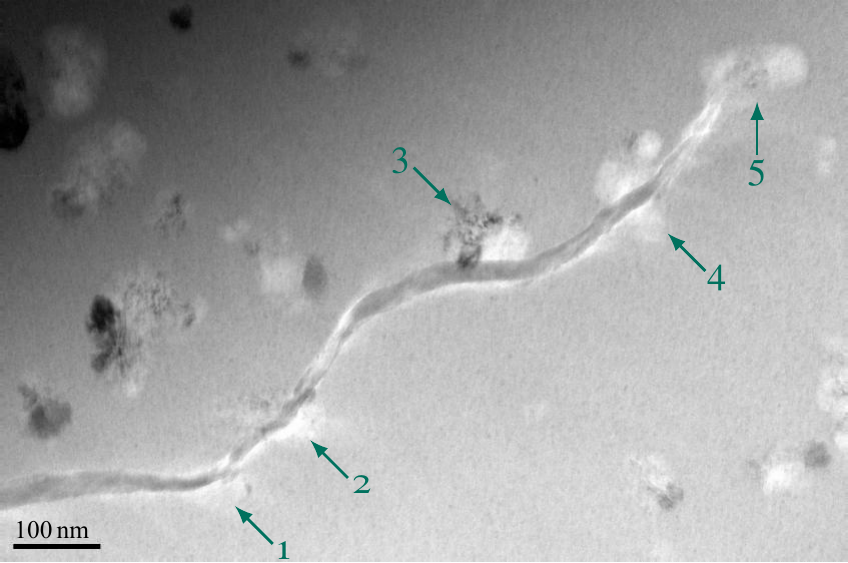}
  \caption{TEM image of a shear band in a Zr-Cu-Ag-Al ribbon after
    deformation. The green arrows mark the positions of crystalline
    precipitates.}
\label{fig:tem-sb-pattern}
\end{figure}

After the DSC heat treatment, the partially crystalline material is
deformed by cold rolling at room temperature in one step to true
strain values of about $\varepsilon = 5\%$ at room temperature,
applying a strain rate of the order of $\dot{\varepsilon} =
1\,\mathrm{s^{-1}}$. Subsequently, specimens for TEM investigations
are prepared by grinding, dimpling, and finally precision ion
polishing (PIPS, Gatan), using a low acceleration voltage of
$2.5\,\mathrm{kV}$ and low incidence angles ($< 4^\circ$) to minimize
damage by the preparation process. The electron transparent samples
are then analyzed in dark-field, bright-field, and high-resolution
transmission electron microscopy modes in a Zeiss Libra 200FE TEM
operated at an acceleration voltage of $200\,\mathrm{kV}$.  Since
intrinsic shear bands formed upon cold rolling could not be found
during the TEM inspections of the thin foil regions, the shear bands
generated as a result of sample preparation and/or subsequent sample
manipulation are studied instead.  Thus, the final sample state
resembles the state in an \textit{in situ} TEM experiment comparable
to the work in Ref.~\onlinecite{Wilde2011}, where the shear bands are
generated at crack tips in the thin TEM foil.

\subsection{Experimental results}

\begin{figure}
  \centering
  \includegraphics{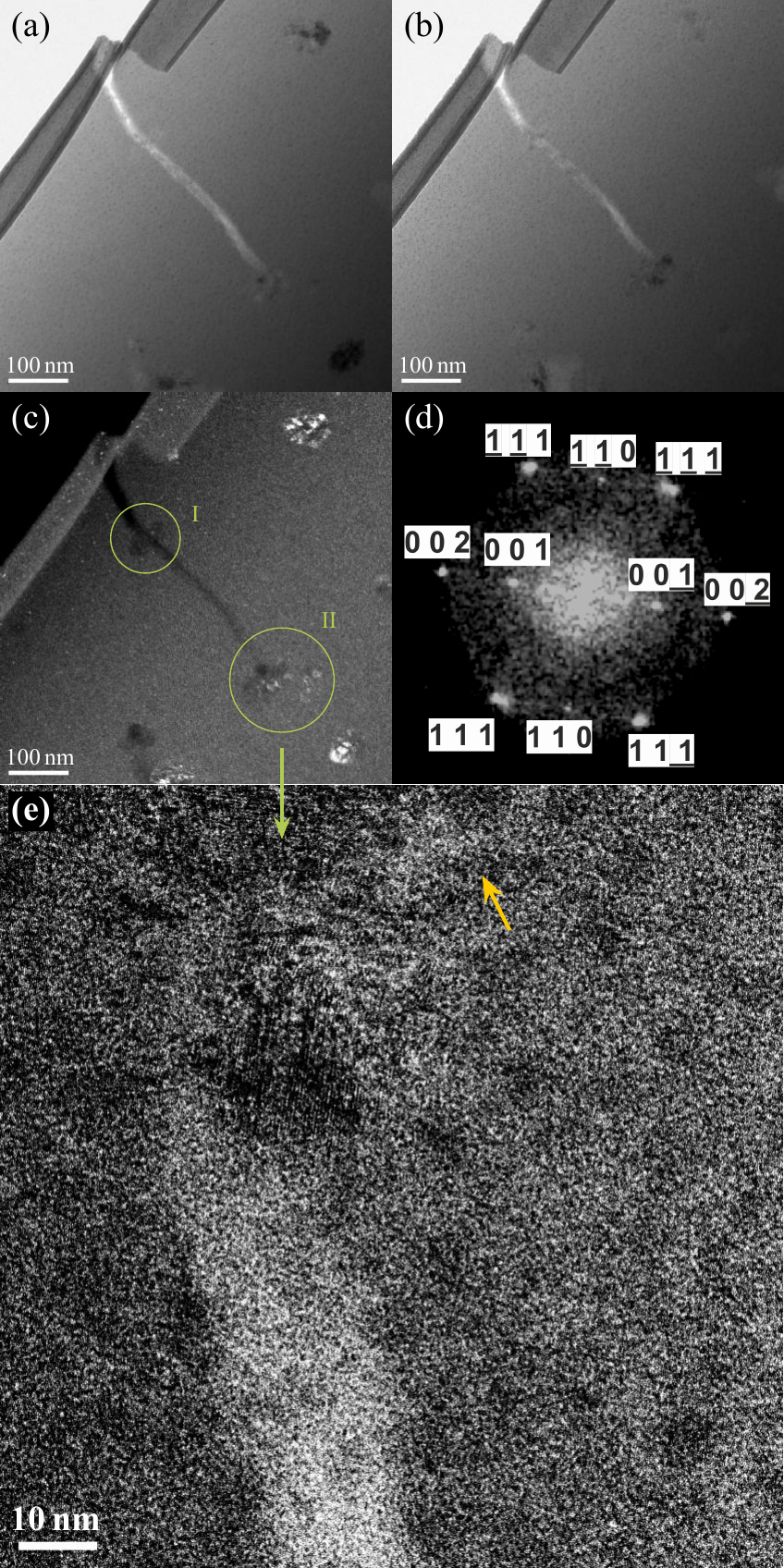}
  \caption{TEM and HRTEM images of a shear band. In the upper left
    corner, the electron transparent hole is visible, which stems from
    the sample preparation for TEM. \mbox{(a--b)} Bright-field images
    of the same area, using a different tilt angle.  The shear band
    appears as a white stripe, with part of it already cracked (bright
    white). (c) Dark-field image of the same area: The
    nanoprecipitates are visible and marked by circles. The crack
    stops at precipitate I; the shear band continues from there to
    precipitate II. (d) The power spectrum of precipitate II along the
    $\langle\overline{1}10\rangle$ zone axis revealing superlattice
    reflections of the martensitic B$19'$ structure.  (e) HRTEM image
    of precipitate II, showing the shear band stopping at the
    precipitate.  The contrast in (c) and (e) is enhanced to improve
    the visibility of the precipitates and the shear band.}
  \label{fig:tem-two-precs-with-sb}
\end{figure}
Figure~\ref{fig:tem-sb-pattern} shows a shear band in a sample after
TEM preparation.  It is noticeable that the shear band switches
propagation directions in the vicinity of precipitates. Between
precipitates 2 and 3, the shear band has an additional bend. The
propagation path, in general, suggests that the shear band is
``attracted'' to the precipitates, possibly due to a stress field
resulting from the density change on crystallization, which explains
the change of direction between precipitates 2 and 3.  Because of the
processing of the samples, it can be excluded that the repeated
bending of the shear band results from a change of the external stress
state: The deformation by cold rolling is performed in a single step,
and the observed shear bands were created during TEM preparation,
resembling an \textit{in situ} experiment.  As the path change in the
presence of precipitates is rather large (in
Fig.~\ref{fig:tem-sb-pattern} around $45^\circ$) and correlated to the
position of the precipitates, it is most likely induced by the
presence of the precipitates.  The literature supports this, as shear
bands in homogeneous metallic glasses (Cu-Zr-based or otherwise) are
straight on the length scale presented
here.\cite{Kumar2007,Wang2008,Roesner2014,Schmidt2015} It is therefore
clear that the precipitates play a major role in influencing
shear-band propagation and thereby the macroscopic plastic deformation
of the material.
TEM images of a second shear band, shown in
Fig.~\ref{fig:tem-two-precs-with-sb}, shed more light on this
interaction. The shear band interacts with two precipitates: A crack
follows the path of the shear band up to the first precipitate and the
shear band continues from the first to the second precipitate.
Because of the visible crack opening near the first precipitate, it
can be ruled out that the shear band originates from the second
precipitate.  While the first precipitate is passed, the second
precipitate, which is encountered centrally, stops the shear band.
This can also be observed at the end of the shear band in
Fig.~\ref{fig:tem-sb-pattern}.  A detailed analysis of the gray-scale
intensity distribution of the high-resolution TEM (HRTEM) image in
Fig.~\ref{fig:tem-two-precs-with-sb}(e) indicates that the shear band
changes its path slightly near the crystalline precipitate but does
not proceed further or shows slip transfer into the precipitate.
Additionally, a smaller, shear-band-like region emerges almost
perpendicular to the previous propagation direction (yellow
arrow). This indicates either a shear-band deflection, or a nucleation
of a new, perpendicular shear band.  Based on the highly local nature
of the intensity distribution and based on the comparison of the
contrast of other precipitates, preparation artifacts can be
excluded. Thus, this type of interaction observed here is part of the
intrinsic interaction mechanism between precipitates and advancing
shear bands.  The power spectrum in
Fig.~\ref{fig:tem-two-precs-with-sb}(d) indicates a precipitate with
B$19'$ crystal structure,\cite{Schryvers1997} which is consistent with
prior observations in Cu-Zr-based MGs.\cite{Pauly2010}

It is clear that the interaction of a propagating shear band with a
distribution of crystalline precipitates depends on the actual stress
state near the shear-band tip, the already-accommodated stress by the
shear-band propagation and the details of the local distribution of
crystalline precipitates, their sizes, and the residual stresses in the
glass matrix due to the formation of the precipitates.  Depending on
these factors, there are various explanations for the observed paths:
The path changes are caused either by a deflection of the shear band,
by the blocking and subsequent renucleation of a new shear band, or by
several nascent shear bands growing together. The unhindered
``passing'' of a precipitate can be explained either by a temporary
path change of the shear band or by the participation of the
precipitate in the plastic deformation.  Because of the necessarily
limited amount of data that can be obtained in the experiment, we
undertake MD simulations to test the aforementioned hypotheses and
investigate their relation to the sample geometry.

\section{Simulation}

\subsection{Simulation setup and analysis}

\begin{figure}[b]
  \centering
  \includegraphics{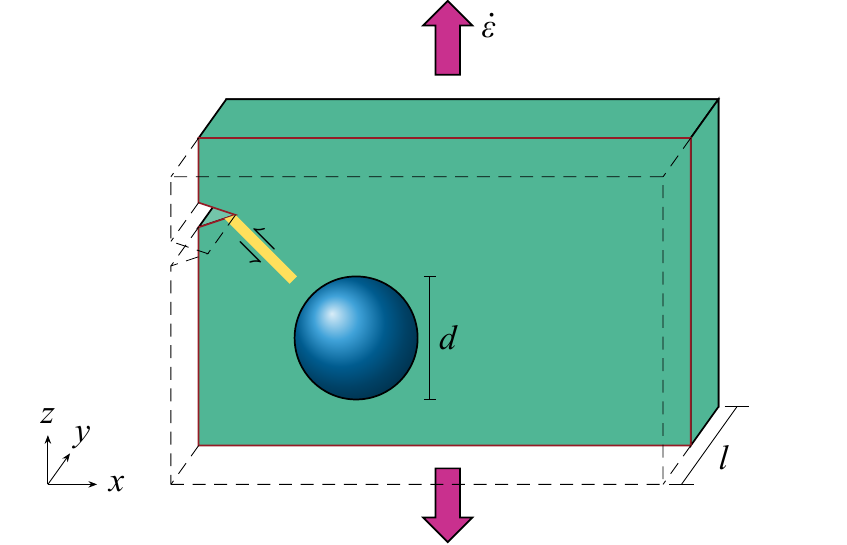}
  \caption{Schematic simulation setup. The picture shows a cut through
    the three-dimensional simulation box in the $xz$ plane at $l/2$. A
    notch is inserted to control the origin of the shear band
    (yellow). The spherical precipitate is shown in blue. The box has
    open boundaries in the $x$ direction and is otherwise periodic. A
    constant strain rate is applied in the $z$ direction.}
  \label{fig:simsetup}
\end{figure}
\begin{figure*}[t!]
  \centering 
  \includegraphics{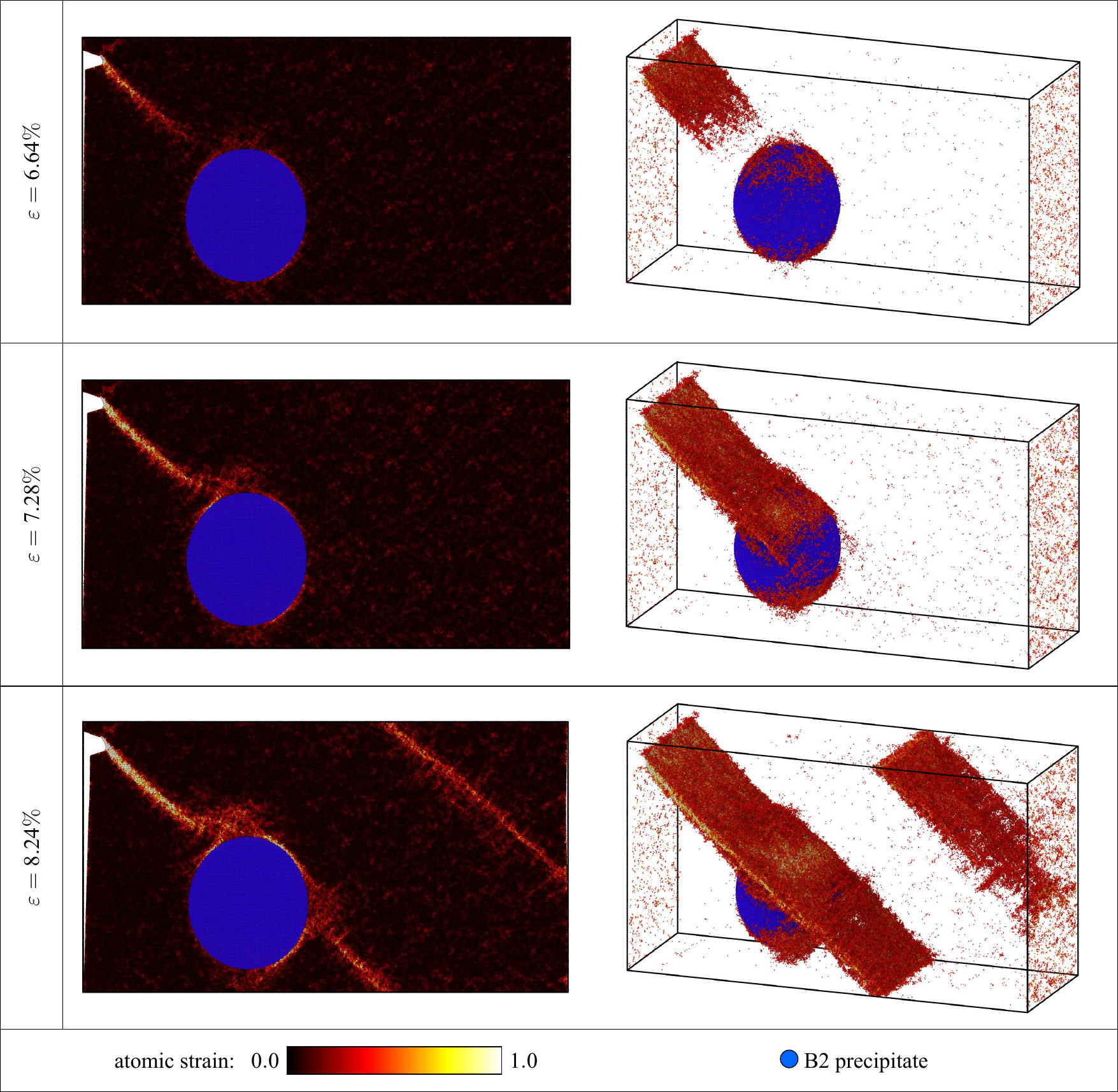}
  \caption[]{Snapshots of a simulation with a $30$-$\mathrm{nm}$
    CuZr precipitate. The shear band wraps around the precipitate and
    continues unhindered.  The glass matrix is colored according to the
    atomic strain. The precipitate atoms are shown in blue if they
    appear in the B2 structure; no defects are visible. The left
    column shows a cut through the middle of the precipitate. On the
    right, all atoms with $\eta_i < 0.3$ are deleted.  A video
    version is provided in Video~\ref{vid:wrapping}.}
  \label{fig:wrapping}
\end{figure*}
\begin{myvideo}
  \centering
  \href{http://link.aps.org/multimedia/10.1103/PhysRevApplied.5.054005}
       {\includegraphics[width=8.6cm]{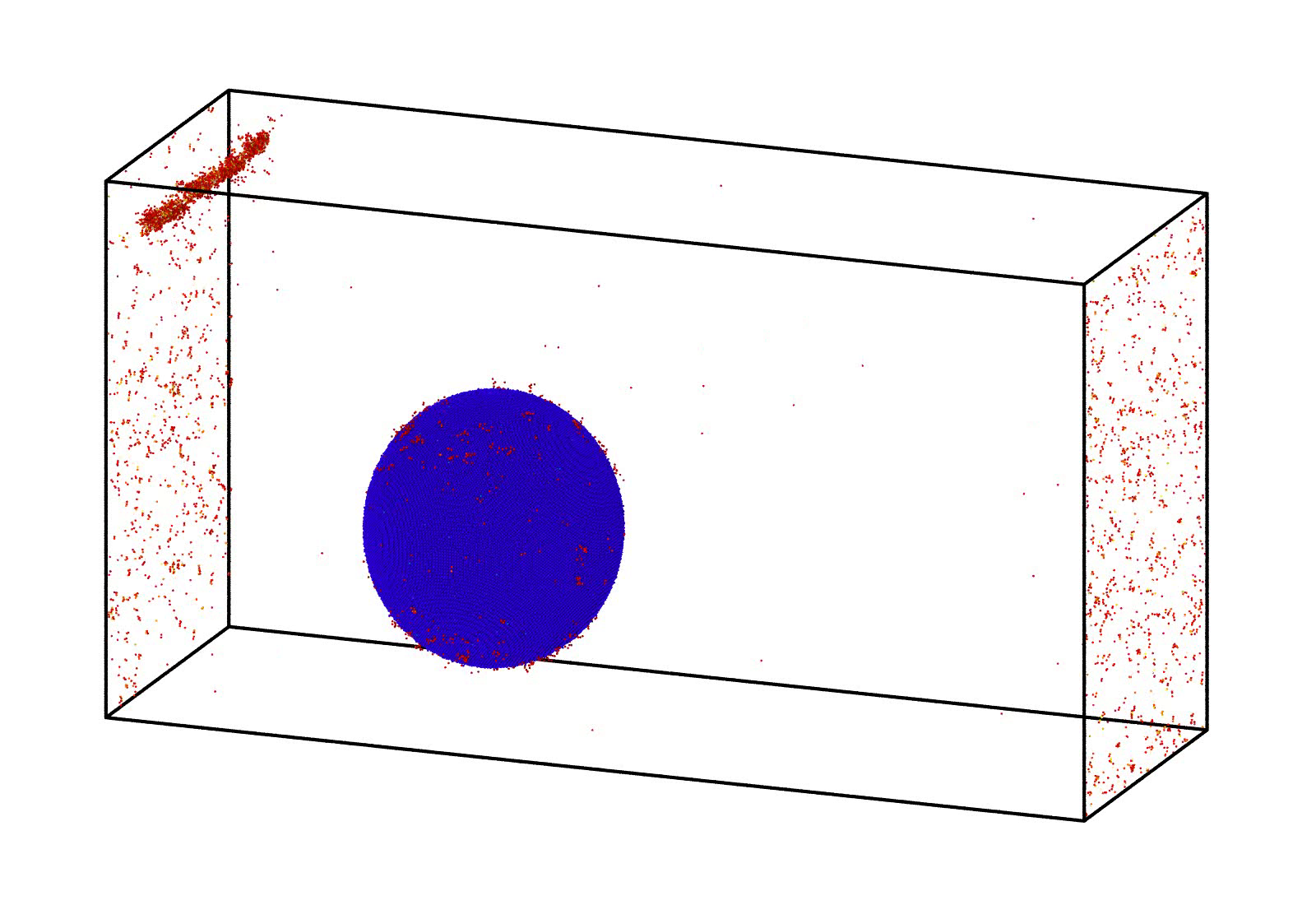}}
  \caption{Simulation of a shear band wrapping around a
    $30$-$\mathrm{nm}$ CuZr precipitate as shown in
    Fig.~\ref{fig:wrapping}.}
  \label{vid:wrapping}
\end{myvideo}

In order to gain more insights into the nanoscale mechanisms of
shear-band interaction with precipitates, we perform a number of MD
computer simulations, which allow for an ``\textit{in situ}''
observation of shear-band propagation.  We use the software
\textsc{lammps} \cite{Plimpton1995} to quench metallic-glass samples,
insert precipitates, and perform mechanical testing. The simulated
metallic glass is a $\mathrm{Cu_{64}Zr_{36}}$ alloy modeled with a
Finnis--Sinclair-type potential by Mendelev \textit{et
  al.}\cite{Mendelev2009} Metallic-glass samples with dimensions $10
\times 10 \times 10\,\mathrm{nm}^3$ and $20 \times 20 \times
20\,\mathrm{nm}^3$ are prepared by melting the material at
$2000\,\mathrm{K}$ and subsequent quenching to $50\,\mathrm{K}$ with a
cooling rate of $0.01\,\mathrm{K/ps}$.

We use the sample geometry illustrated in Fig.~\ref{fig:simsetup}
to investigate the influence of preexisting precipitates on an
approaching shear band.  Size effects are studied by varying the
diameter $d$ of the precipitates and by controlling the distance
between periodic precipitate images by varying the box width $l$.
Sample sizes are $120\,\mathrm{nm} \times l \times
60\,\mathrm{nm}$, with $l = 10\,\mathrm{nm}$, $20\,\mathrm{nm}$,
$30\,\mathrm{nm}$, and $40\,\mathrm{nm}$.  With a single precipitate
per simulation box, this corresponds to number densities for the
precipitates of $1.4\times10^{22}/\mathrm{m^3}$,
$6.9\times10^{21}/\mathrm{m^3}$, $4.6\times10^{21}/\mathrm{m^3}$, and
$3.5\times10^{21}/\mathrm{m^3}$, respectively.  The Cu-Zr glass
samples are replicated to reach the desired box dimensions, and
spherical precipitates with diameters from $3\,\mathrm{nm}$ to
$40\,\mathrm{nm}$ were inserted.  For this, a hole is cut into the
glass matrix with a size chosen to accommodate the precipitate without
overlapping atoms.  For the CuZr precipitate, we use the
experimentally observed B2 structure.\cite{Pauly2007, Sun2005,
  Das2007, Jiang2007} The B$19'$ structure, which was found in the
precipitates in the experimental part of this paper, is a distortion
of the B2 structure.\cite{Schryvers1997} A notch controls the origin
of the shear band and makes sure that it always hits the precipitate.
Periodic boundary conditions are applied in the $y$ and $z$ directions
and open boundaries in the $x$ direction. The resulting structure is
equilibrated at $50\,\mathrm{K}$ for $2\,\mathrm{ns}$ with a barostat
at ambient pressure in periodic directions. After equilibration, no
long ranging stress field around the precipitate is left; any
mismatches are accommodated by the glass during the interface
creation.

The resulting composite samples are deformed at $50\,\mathrm{K}$
under a constant tensile strain rate of $\dot{\varepsilon} = 4 \times
10^7 / \mathrm{s}$ in the $z$ direction up to a total strain of at least
$10\%$.  The trajectories from equilibrated to fully deformed samples
are analyzed to observe the shear-band propagation path.  The shear
band is identified using the von Mises local shear invariant $\eta_i$
\cite{Shimizu2007} as implemented in the visualization tool
\textsc{Ovito}.\cite{Stukowski2010} Atoms with a local shear greater
than around $0.2$ are assigned to the shear band.  To observe plastic
deformation events in the crystalline phase, we perform atomic
structure identification, which can identify crystal structures,
stacking faults, and other defects.\cite{Stukowski2012}

\subsection{Wrapping and blocking}

\begin{figure}
  \centering
  \includegraphics{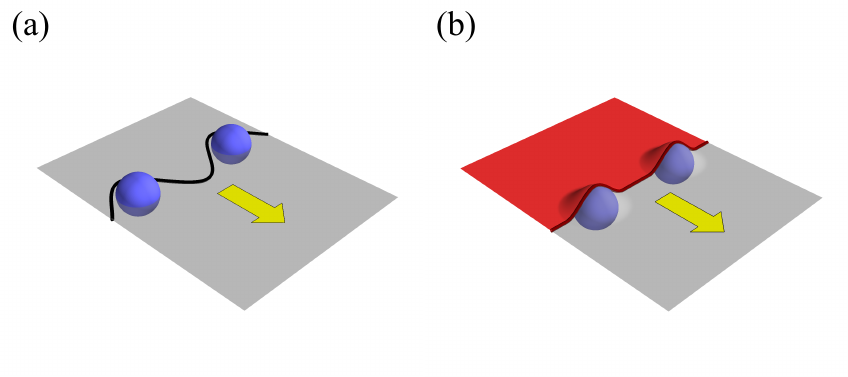}
  \caption{Comparison of the Orowan mechanism (a) with a shear band
    wrapping around precipitates (b). The slip plane is shown in gray,
    the precipitates in blue, the dislocation as a black line, and the
    shear band as a red plane. The precipitates can stop a dislocation
    because it must remain in a defined slip plane, while the shear
    band can temporarily leave its slip plane and continue unhindered
    afterwards.}
  \label{fig:sb-orowan}
\end{figure}

The first observed interaction mechanism between a propagating shear
band and a preexisting precipitate is shown in
Fig.~\ref{fig:wrapping} and Video~\ref{vid:wrapping}. Here, the shear
band wraps around the precipitate like a carpet moving over a small
obstacle.  The particle does not deform and simply moves along with
one half of the glass matrix.  We call this mechanism the
\emph{wrapping mechanism}. This kind of athermal mechanism is
virtually unknown in crystalline materials. The closest analog in a
crystal---dislocation climb---is purely thermally activated. It is
therefore instructive to compare the two material classes, as shown in
Fig.~\ref{fig:sb-orowan}. In a crystalline material, a dislocation
moves on defined slip planes. A change of slip plane is possible only
for the screw components of the dislocation,\cite{Cottrell1953} is
connected with a high energy barrier, and is usually observed only in
stage~III work hardening.\cite{Hirth1982} Therefore, the Orowan
mechanism applies: The dislocation is bent around the obstacle and
finally forms dislocation rings
[Fig.~\ref{fig:sb-orowan}(a)].\cite{Gottstein2004} These rings may
pile up, thereby hardening the material. In a metallic glass, as in
any isotropic material, all slip directions are equivalent. Only an
applied external stress differentiates the directions. Under tensile
stress, the planes oriented in $45^\circ$ angles towards the tensile
axis experience the highest resolved stress. An obstacle can be
avoided simply by temporarily and locally changing the slip path,
thereby wrapping around the obstacle [Fig.~\ref{fig:sb-orowan}(b)].
Depending on the precipitate distance, this wrapping mechanism is
observed for precipitates with diameters smaller than
$25\,\mathrm{nm}$ to $35\,\mathrm{nm}$ in our simulations.

\begin{figure*}[t!]
  \centering 
  \includegraphics{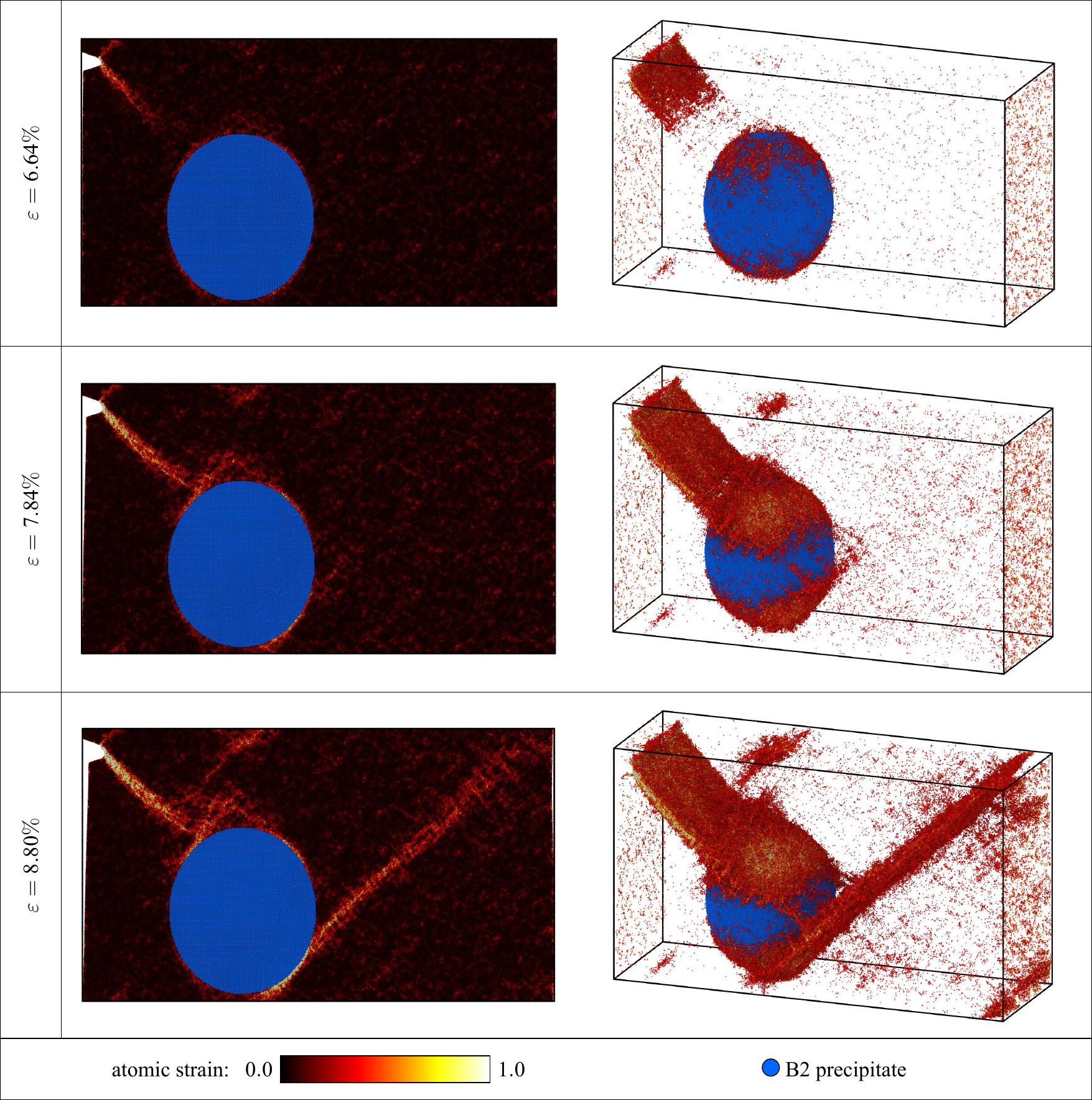}
  \caption{Snapshots of a simulation with a $37.5$-$\mathrm{nm}$
    CuZr-precipitate. The shear band is blocked by the precipitate,
    while a second shear band is immediately nucleated in another
    plane of high resolved shear stress.  The glass matrix is colored
    according to the atomic strain. The precipitate atoms are shown in
    blue if they appear in the B2 structure; no defects are
    visible. The left column shows a cut through the middle of the
    precipitate. On the right, all atoms with $\eta_i < 0.3$ are
    deleted.  A video version is provided in Video~\ref{vid:blocking}.}
  \label{fig:blocking}
\end{figure*}
\begin{myvideo}
  \centering
  \href{http://link.aps.org/multimedia/10.1103/PhysRevApplied.5.054005}
       {\includegraphics[width=8.6cm]{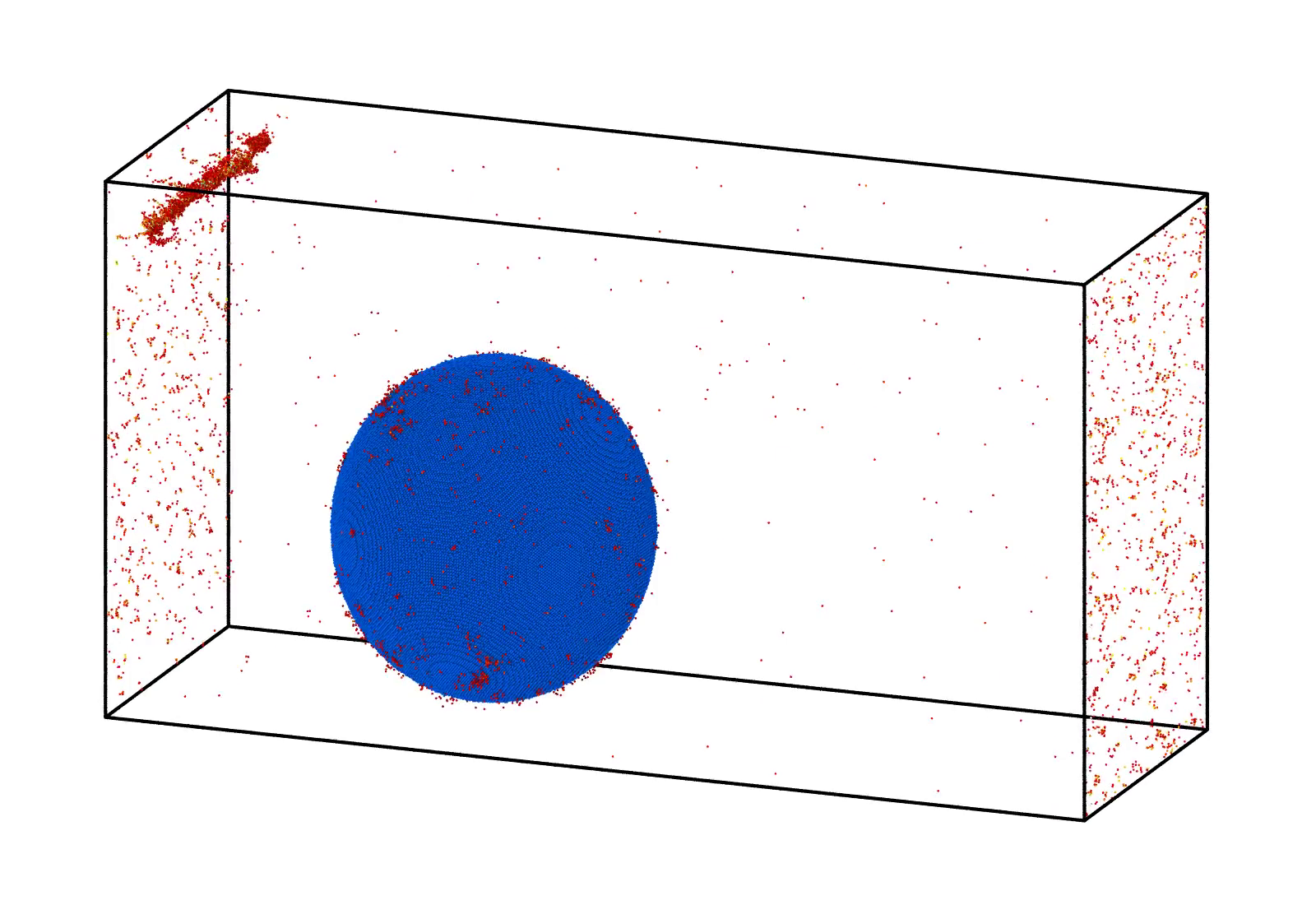}}
  \caption{Simulation of a shear band being blocked by a
    $37.5\,\mathrm{nm}$ CuZr precipitate as shown in
    Fig.~\ref{fig:blocking}.}
  \label{vid:blocking}
\end{myvideo}

An alternative mechanism appears for increasing diameters and
decreasing distances between precipitates: The shear band is
\emph{blocked} by the precipitate.  This causes the simultaneous
nucleation of a second shear band perpendicular to the first one on
the opposite side of the precipitate as shown in
Fig.~\ref{fig:blocking} and in Video~\ref{vid:blocking}. Again, there
is no slip transfer to the crystalline phase.  The paths of the shear
bands in this simulation and in the HRTEM image of the experimental
sample [Fig.~\ref{fig:tem-two-precs-with-sb}(e)] are comparable.  Both
show what looks like a shear band that starts to wrap around the
precipitate but does not propagate further.  In the experiment,
though, no fully formed shear band appears perpendicular to the
original one.  This may be a result of the more complex stress state
or the fact that a new shear band can nucleate at another precipitate
that is not visible in the images. Still, a region resembling a
nascent shear band appears perpendicular to the original propagation
direction, strengthening the agreement with the simulation results.

\begin{figure}
  \centering
  \includegraphics{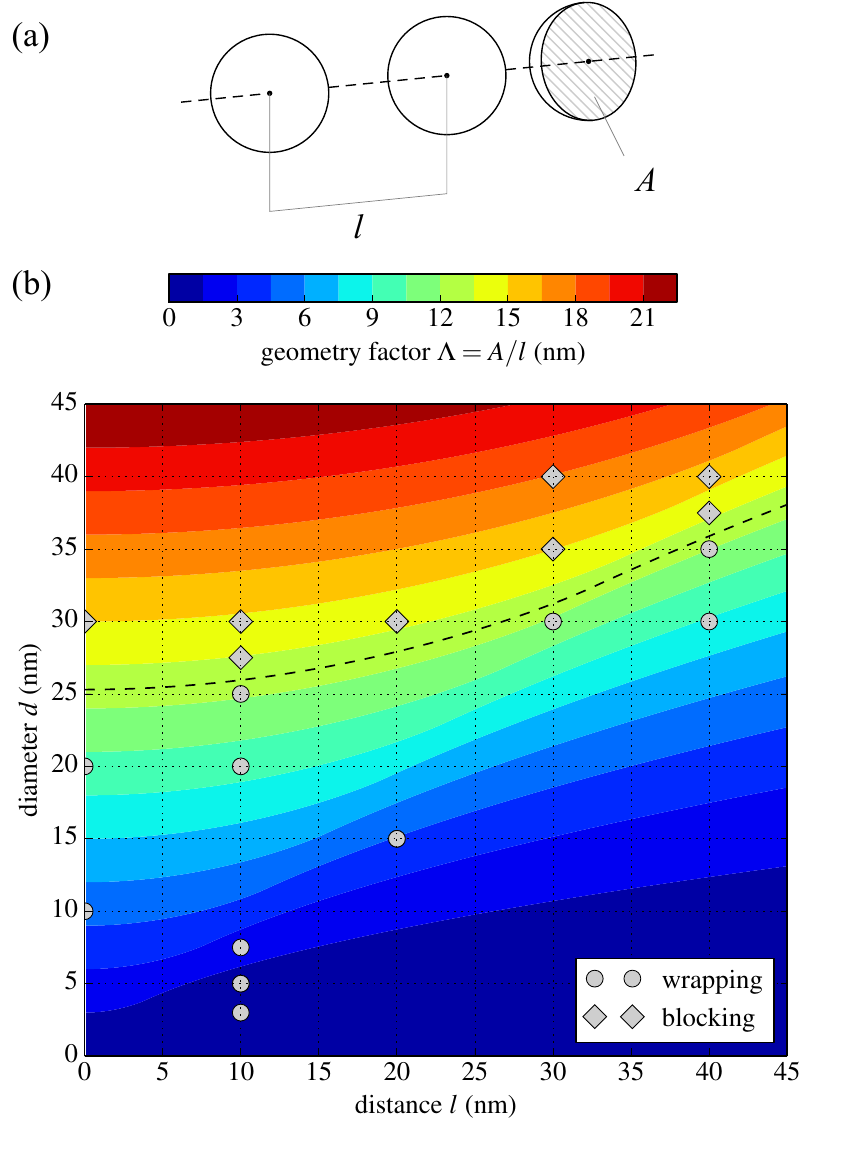}
  \caption{Influence of the precipitate size and distance on the
    mechanism. (a) Explanation of the parameter $l$, distance of the
    precipitates, and $A$, the area of a cut through the
    precipitate. (b) shows a contour plot of the empirical
    geometry factor $\Lambda = A/l$. The dashed line shows the
    critical value of $\Lambda$ for a transition from the wrapping to
    the blocking mechanism. Data points are MD simulations. The data
    points at $l=0\,\mathrm{nm}$ are infinite cylinders along the
    $y$ axis, representing the case of overlapping spheres with
    infinitesimally small distances between their centers.}
  \label{fig:geometry}
\end{figure}

A systematic investigation of the parameters of precipitate distance
and size reveal a clear correlation with the mechanism.  For a more
quantitative analysis, we define an empirical parameter $\Lambda$,
which is given by the ratio of the cross-sectional area $A$ of the
precipitate divided by the distance of precipitate centers $l$ as
shown in Fig.~\ref{fig:geometry}(a).  Similar to the derivation of the
Orowan stress, the crystalline volume fraction $f =
V_\mathrm{precipitates} / V$ can be estimated from the average
crystallite distance by the relation\cite{Gottstein2004}
\begin{equation}
  l = \frac{d/2}{\sqrt{f}}
  \quad \Leftrightarrow \quad
  f = \frac{d^2}{4l^2}.
\end{equation}
Using that, we can express $\Lambda$ only in terms of volume fraction
and precipitate geometry:
\begin{equation}
  \Lambda = \frac{A}{l} = \frac{A \sqrt{f}}{d/2}.
\end{equation}
For nonoverlapping precipitates ($d < l$), it is $A = \pi d^2/4$ and
$\Lambda$ reduces to
\begin{equation}
  \Lambda = \pi \frac{d}{2} \sqrt{f}.
\end{equation}
When the precipitates overlap ($d > l$), we reduce the area $A$ to
remove the overlapping circle segments. In the limit $l \rightarrow
0$, $\Lambda$ corresponds to $A/l$ of an infinite cylinder parallel to
the shear-band front. Figure~\ref{fig:geometry}(b) shows a contour
plot of the $\Lambda$ parameter as a function of the precipitate diameter
and distance. The data points in the plot represent results from our
MD simulations, divided into those showing the wrapping and those
showing the blocking mechanism.  For the simulation geometry used in
this work, there exists a given $\Lambda$ that clearly separates the
two mechanisms:
\begin{equation}
  \Lambda_\mathrm{crit} \approx 12.65\,\mathrm{nm}
\end{equation}
This $\Lambda_\mathrm{crit}$ is not universal, as, for example, the
distance between precipitate and notch is not varied.  A test
simulation finds that increasing the distance from the notch and therefore
increasing shear-band length favors the wrapping mechanism.

\begin{figure}
  \centering
  \includegraphics{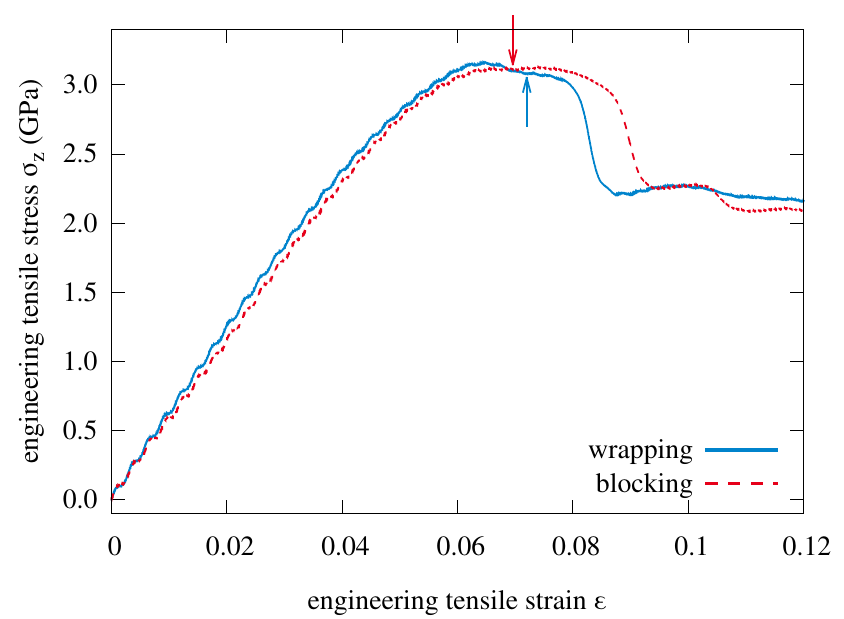}
  \caption{Stress-strain curves of samples which exhibit the wrapping
    or blocking mechanisms. The arrows indicate when the shear band
    hits the precipitate.}
\label{fig:stress-strain-B2}
\end{figure}
\begin{figure}[b]
  \centering
  \includegraphics{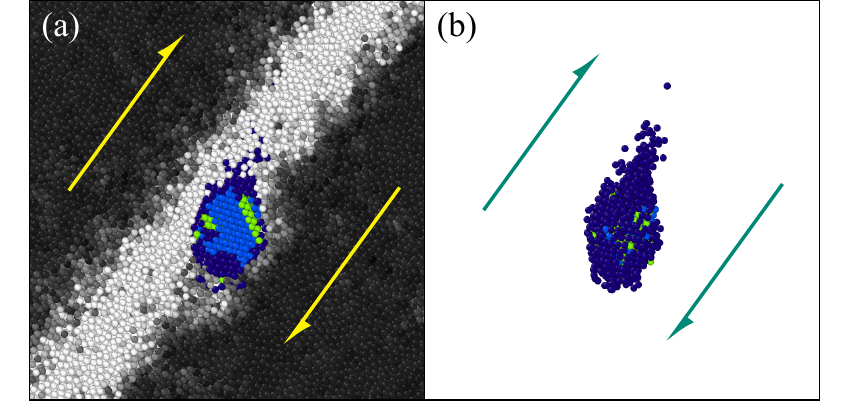}
  \caption{Mechanical dissolution of a $3$-$\mathrm{nm}$ copper
    particle in a shear band.  The gray-scale color coding shows
    atomic strain $\eta_i$ from $0.0$ (black) to $1.0$ (white). The
    precipitate is shown in color: Light blue atoms are in the fcc
    structure, green atoms are in a stacking fault, and dark blue
    atoms are disordered.  The arrows indicate the shear direction.
    Subfigure (a) shows a cut through the middle of the
    nanoprecipitate, while (b) shows only the atoms that initially
    belonged to the nanoprecipitate.}
  \label{fig:dissolve}
\end{figure}

Using these formulas, we can also estimate $\Lambda$ for the
experimental results. Given a number density of precipitates $n =
10^{20}\,\mathrm{m^{-3}}$ and particle diameters of around
$70\,\mathrm{nm}$, we obtain
\begin{align}
  f &= \frac{V_\mathrm{precipitates}}{V}
    = n \frac{4}{3} \pi (35\,\mathrm{nm})^3
    \approx 1.8\% \\ 
  \Lambda_\mathrm{exp} &= \pi \times 35\,\mathrm{nm} \times \sqrt{1.8\%}
                     \approx 14.7\,\mathrm{nm} > \Lambda_\mathrm{crit}.
\end{align}
This is consistent with the fact that the shear bands that are
observed in the glass samples are blocked by the precipitates. Still,
the value is close to $\Lambda_\mathrm{crit}$, which means that
slightly smaller precipitates (or precipitates that are not hit
centrally, virtually decreasing the prefactor $d$) may be susceptible
to wrapping.

\begin{figure*}
  \centering
  \includegraphics{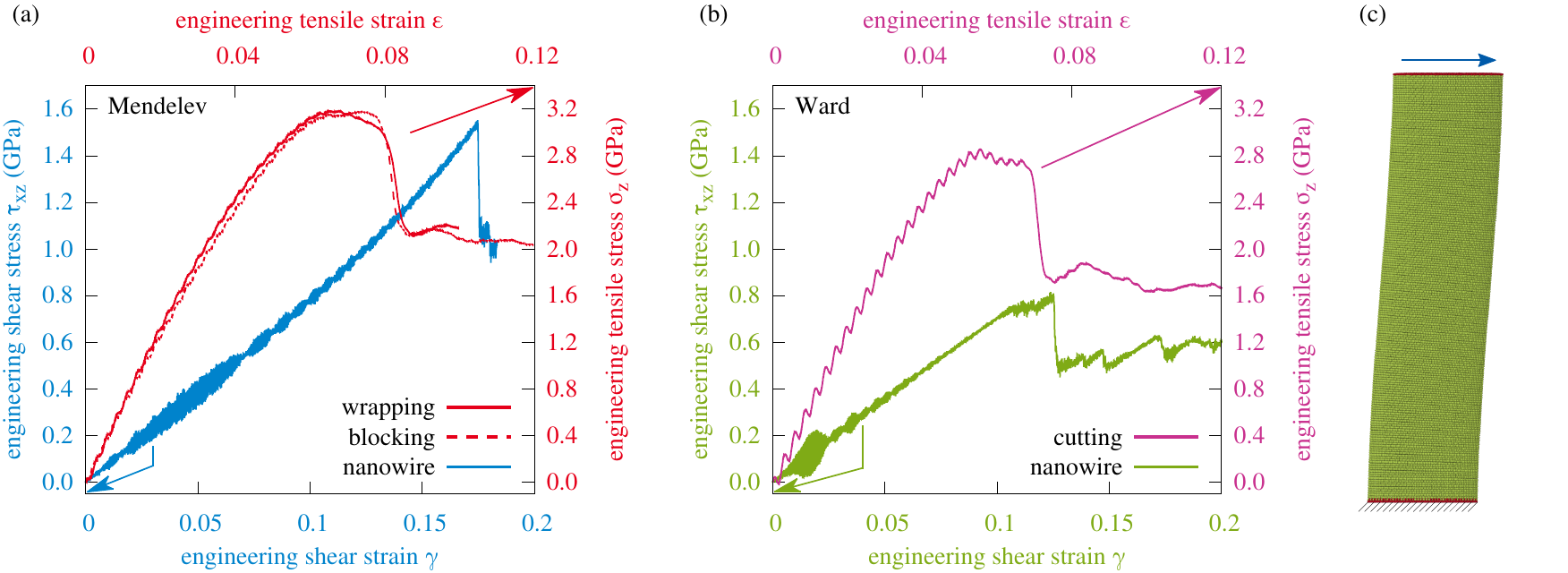}
  \caption{Stress-strain curves for the composites with fcc copper
    precipitates and for a $10$-$\mathrm{nm}$ copper nanowire.  (a)
    Curves from simulations with the Mendelev potential. The copper
    precipitates do not deform plastically and show similar behavior
    to the CuZr precipitates discussed earlier.  The curve of the
    copper nanowire explains this: The yield stress, and therefore the
    critical shear stress, is higher than the highest resolved shear
    stress in the steady state in the composite.  To be able to
    compare the tensile stress in the composite with the shear stress
    in the nanowire, the tensile stress axis is scaled with a factor
    $0.5$ corresponding to the Schmid factor for the plane of highest
    resolved shear stress.  (b) Curves from simulations with the Ward
    potential. Now the critical stress for heterogeneous dislocation
    nucleation is lower than the steady-state stress in the composite
    and the precipitate deforms plastically.  (c) The simulation setup
    for the nanowire.  The wire is sheared in the $[110]$ direction of the
    fcc crystal structure on the $(111)$ plane.  The red atoms are
    fixed and the atoms on the top are shifted with a constant
    velocity to shear the nanowire.}
  \label{fig:nanowire}
\end{figure*}

Figure~\ref{fig:stress-strain-B2} shows examples of the stress-strain
curves of samples that exhibit either the wrapping or the blocking
mechanism.  Neither a pronounced ductility nor significant
strain hardening can be observed.  This lack of strain hardening is
in accordance with experimental data for tensile tests on Cu-Zr-based
metallic glasses with crystalline
precipitates.\cite{Pauly2010,Barekar2010,Pauly2010a} It also fits with
a recent study of compression tests of Cu-Zr-based metallic glass,
that finds an effect of particle size on mechanical parameters but no
particle hardening.\cite{Wang2014a}

\subsection{Plastic deformation of the crystalline phase}

\begin{figure*}[p]
  \centering 
  \vspace{-1cm}
  \makebox[\textwidth][c]{%
    \includegraphics{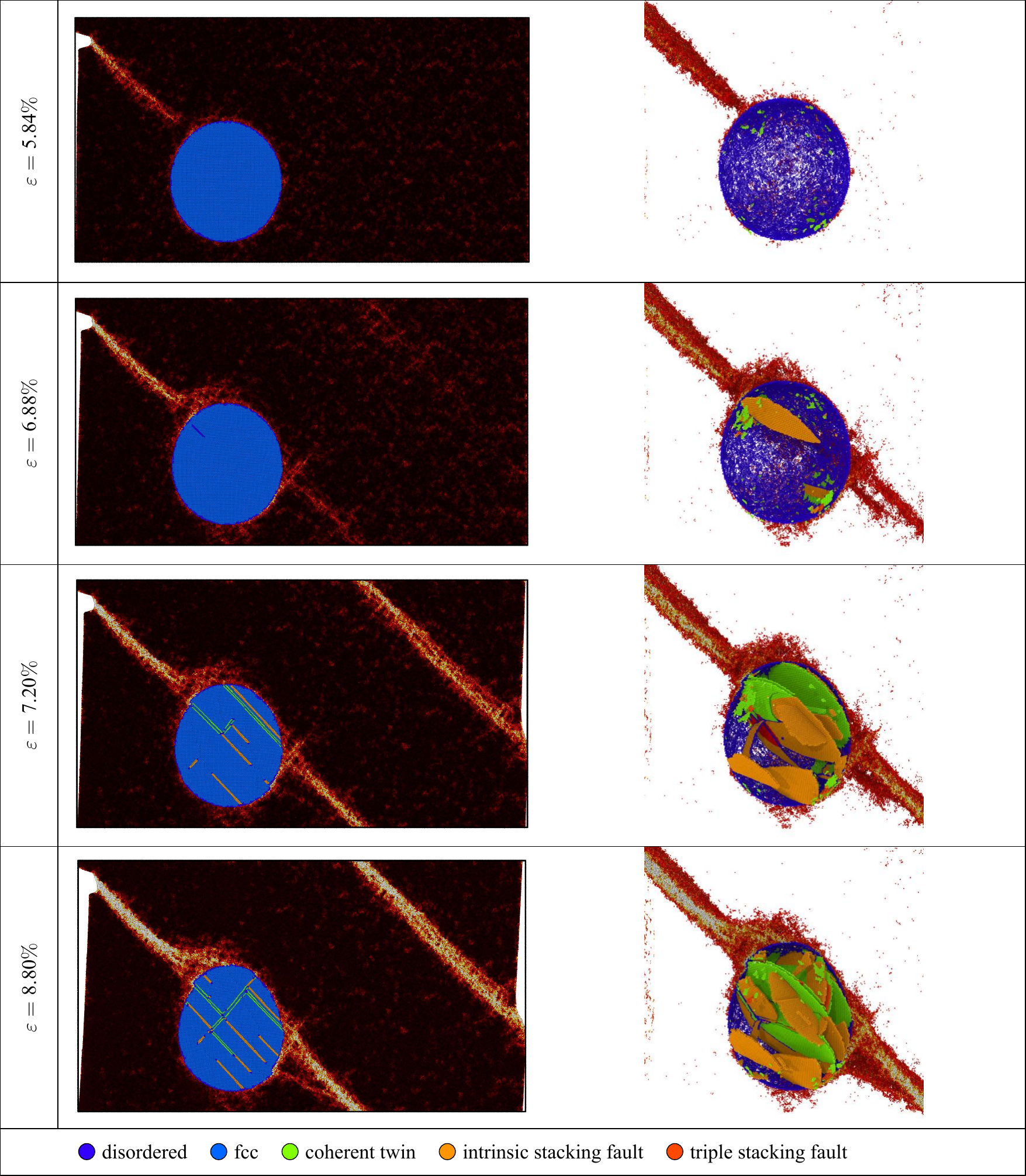}%
  }
  \caption{Snapshots of a simulation with a $30$-$\mathrm{nm}$ copper
    precipitate. This simulation uses the Ward potential, in which the
    critical stress for dislocation nucleation is realistic. As the
    precipitate is sufficiently soft, the shear band can cut through
    it. The glass matrix is colored according to the atomic strain
    (the same scale as Figs.~\ref{fig:wrapping} and
    \ref{fig:blocking}). Defects in the fcc crystal structure are
    colored according to the legend. The left column shows a cut
    through the middle of the precipitate. On the right, all atoms
    with $\eta_i < 0.3$ and all fcc-coordinated atoms are deleted.  A
    video version is provided in Video~\ref{vid:plastic-precipitate}.}
  \label{fig:plastic-precipitate}
\end{figure*}
\begin{myvideo}
  \centering
  \href{http://link.aps.org/multimedia/10.1103/PhysRevApplied.5.054005}
       {\includegraphics[width=8.6cm]{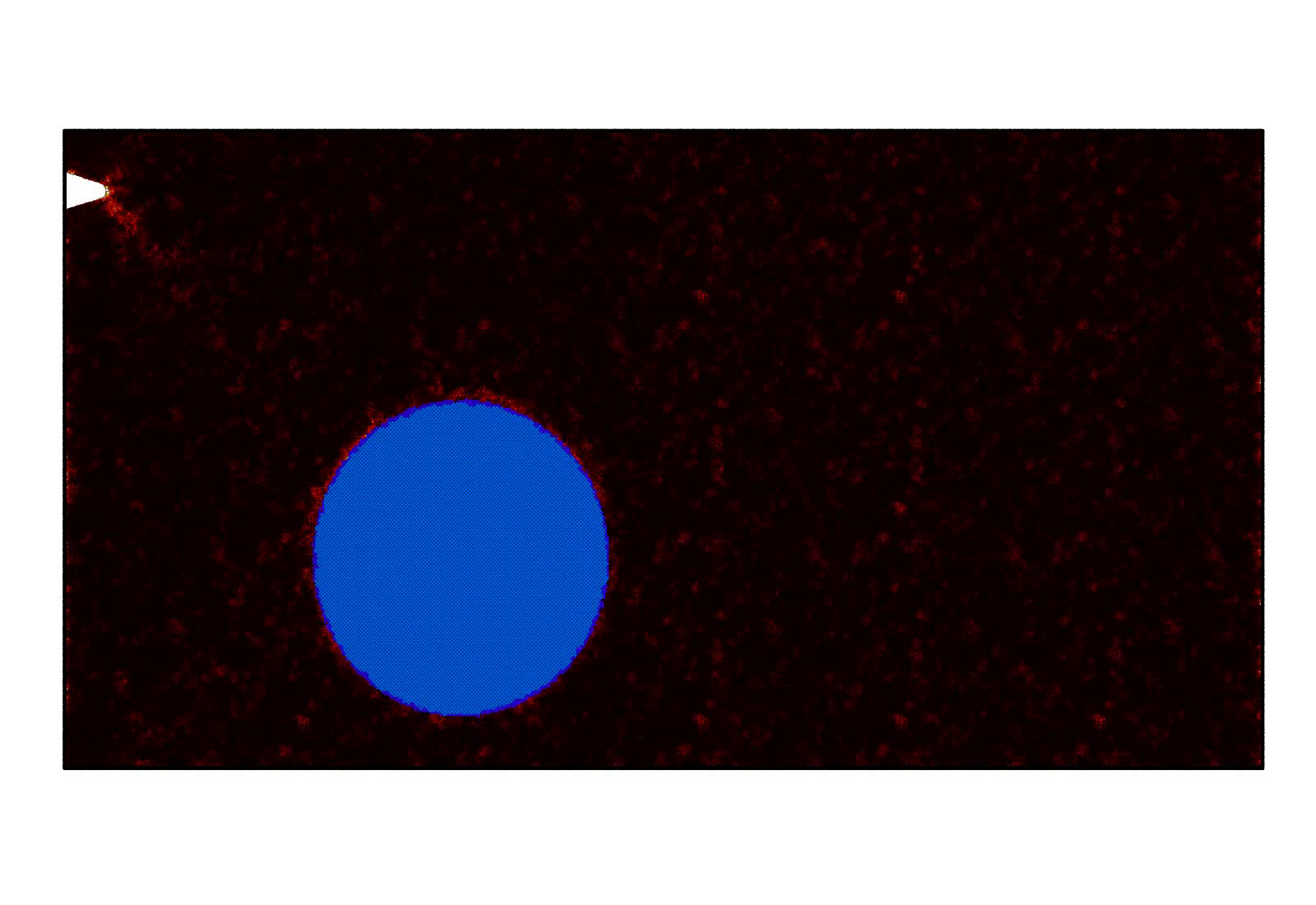}}
  \caption{Simulation of a shear band cutting a $30$-$\mathrm{nm}$
    copper precipitate as shown in
    Fig.~\ref{fig:plastic-precipitate}.}
  \label{vid:plastic-precipitate}
\end{myvideo}
The precipitates discussed until now were all ``hard,'' i.e.,
not susceptible to plastic deformation under the simulation
conditions: Because of the high antiphase-boundary energy in the B2
structure, superdislocations or twinning with respect to a
martensitic transformation would be needed for a plastic deformation
of the precipitates.  The stress available at the shear-band tip is
not sufficient to nucleate these defects, which explains why no
plastic deformation of the crystalline phase can be observed in our
simulations.  In experiments, deformation-grown precipitates show
twinning defects which may cause softening effects and make the
precipitate susceptible to plastic deformation.\cite{Cao2007} As a
model for a softer precipitate, we thus exchange the B2 crystal phase
for fcc copper.

For precipitates that are small relative to the shear-band width, the
nanocrystals undergo mechanical dissolution.\cite{Lund2007,Albe2013}
This is also observed in our setup with $3$-$\mathrm{nm}$ particles as
shown in Fig.~\ref{fig:dissolve}.  In samples with larger diameters,
the precipitates do not deform plastically but instead show the same
wrapping and blocking interactions as described before for the
``hard'' precipitates.  Even if the $(111)$ glide plane is oriented
parallel to the shear-band direction to maximize resolved shear stress
on the preferred fcc slip plane, we do not observe slip transfer into
the nanoprecipitate.  The corresponding stress-strain curves are shown
in Fig.~\ref{fig:nanowire}(a).  To explain this, we estimate the
critical stress for heterogeneous dislocation nucleation in fcc copper
by shearing a nanowire on the $(111)$ plane in the $[110]$ direction [see
Fig.~\ref{fig:nanowire}(c) for the simulation setup]. For this, we
hold the lower layer of atoms fixed and move the top layer of atoms
with a constant velocity in the $[110]$ direction to achieve
volume-conserving shear.  The diameter of the nanowire is
$10\,\mathrm{nm}$, i.e., on the order of the smaller
precipitates to maximize surface effects.  The resulting shear stress
over shear curve is also plotted in Fig.~\ref{fig:nanowire}(a). The
yield stress of the nanowire, $\tau_\mathrm{crit}$, is an estimate for
the stress needed for heterogeneous dislocation nucleation at the
glass-precipitate interface.  For comparing the shear stress in the
nanowire with the tensile stress in the composite, we assume a Schmid
factor of $0.5$ and scale the tensile axis by a factor $0.5$ compared
to the shear stress axis.  This graphically estimates an upper bound
of the shear stress $\tau_m$ that arrives at the precipitate,
corresponding values are given in Table~\ref{tab:yield-data}.  The
upper bound of the estimated maximum resolved shear stress in the
composite is comparable to the lower bound for heterogeneous
dislocation nucleation.  While this suggests that dislocation
nucleation may be possible, it is important to keep in mind that the
nanowire shear test provides only a lower bound in an idealized case
and that the actual shear stress available to nucleate a dislocation may be
lower than $0.5\sigma_z$, which is why no plastic deformation of the
precipitate is observed.  Evidently, the value of
$\tau_\mathrm{crit}$ in the Mendelev potential is much too high, and
therefore the Cu precipitates are ``harder'' than expected, which is a
deficiency of the potential model for pure Cu.
\begin{table}
  \caption{Comparison of maximum stress $\sigma_m$ and the
    corresponding resolved shear stress $\tau_m$ with the critical
    shear stress $\tau_\text{crit}$ for heterogeneous dislocation
    nucleation in copper in the Mendelev and Ward potentials. The
    calculation of the resolved shear stress assumes a Schmid factor
    of $0.5$.}
  \label{tab:yield-data}
  \centering
  \vspace{0.2cm}
  \begin{ruledtabular}
  \newcolumntype{d}[1]{D{.}{.}{#1}}
  \begin{tabular}{ld{1.1}d{1.2}d{1.2}}
    & \multicolumn{1}{c}{$\sigma_m$ (GPa)}
    & \multicolumn{1}{c}{$\tau_m$ (GPa)}
    & \multicolumn{1}{c}{$\tau_\text{crit}$ (GPa)} \\
    \noalign{\vskip 0.075cm}
    \colrule
    \noalign{\vskip 0.075cm}
    Mendelev       & 3.2      & 1.6   & 1.55 \\
    Ward           & 2.9      & 1.45  & 0.75 \\
  \end{tabular}
  \end{ruledtabular}
\end{table}

Because of this, we switch to a different potential, which provides
a better description of crystalline Cu.  The Finnis--Sinclair-type
potential by Ward \textit{et al.}\ \cite{Ward2012preprint} is created
by using preexisting potentials for the elemental phases
\cite{Zhou2004} and fitting the cross terms to the intermetallic
phases.  As shown in the \hyperref[sec:GSF]{Appendix}, this potential
has a more realistic unstable stacking-fault energy and critical
stress for homogeneous dislocation nucleation than the Mendelev
potential.  As shown in Fig.~\ref{fig:nanowire}(b) and
Table~\ref{tab:yield-data}, the critical stress for heterogeneous
nucleation is much lower than even the steady-state stress in the
composite, easily allowing plastic deformation of the particle.

The results for a $30$-$\mathrm{nm}$ copper precipitate are shown in
Fig.~\ref{fig:plastic-precipitate} and in
Video~\ref{vid:plastic-precipitate}. The precipitate is \emph{cut} by
the shear band, and slip transfer through the particle can be observed.
This mechanism replaces the previously discussed blocking of the shear
band if the nanoprecipitates are ``soft'': For the crystal to partake
in the plastic deformation, dislocation nucleation must be possible at
shear stresses below the highest resolved shear stress in the metallic
glass at yield.  Despite the participation of the crystalline phase in
the plastic deformation, the stress-strain curve in
Fig.~\ref{fig:nanowire}(b) shows the distinctive stress drop connected
with a single critical shear band and no strain hardening.  The reason
is that in this setup the crystalline phase accounts only for roughly
$5\,\mathrm{vol\%}$ of the sample. This means that the macroscopic
mechanical properties are still dominated by the metallic glass.  As
the shear band can simply cut through the crystal, the precipitate
poses no obstacle to the percolation of the critical shear band.  For
a larger crystalline volume, a ductile crystalline phase could
possibly also constrain the shear bands.\cite{Hofmann2008}

\section{Discussion}
\label{sec:discussion}

\begin{figure}
  \centering
  \includegraphics[]{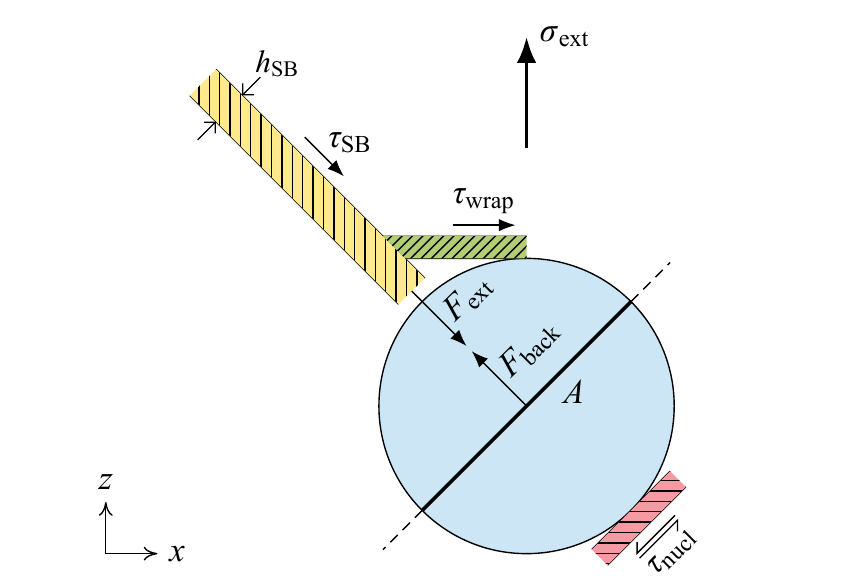}%
  \caption{Projection of the forces and stresses acting around the
    precipitate (blue circle) onto the $xz$ plane.  Shear bands are
    shown as hatched areas, where yellow signifies the arriving
    shear band, green the path for wrapping, and red the site for the
    nucleation of a new shear band.}
  \label{fig:model-geometry}
\end{figure}

Using TEM imaging, we observe shear-band bending around or close to
precipitates, an attraction of shear bands to the precipitates, and
shear bands being blocked by precipitates.
In our MD simulations, we find four mechanisms of interaction between
shear bands and precipitates:
\begin{enumerate}
  \renewcommand{\labelenumi}{\theenumi}
  \renewcommand{\theenumi}{(\roman{enumi})}
\item precipitates that are small relative to the shear-band width
  dissolve mechanically,
\item shear bands can wrap around precipitates,
\item shear bands are blocked by precipitates, and
\item shear bands cut through precipitates, and slip transfer into
  the crystalline phase takes place.
\end{enumerate}
Which of these mechanisms is active for a given precipitate depends on
the competition between the propagation of the existing shear band,
the heterogeneous nucleation of a new shear band, and the
heterogeneous dislocation nucleation in the precipitate.
The wrapping-to-blocking transition can be quantified by the parameter
$\Lambda = A / l \propto A \sqrt{f} / d$.  Below $\Lambda =
\Lambda_\mathrm{crit}$, the wrapping mechanism is favored.  This value
can be be explained by the following simple argument. When the shear
band reaches a precipitate which does not deform, the shearing of the
sample momentarily stops.  The stress $\tau_\mathrm{SB}$ in the shear
band resulting from the externally applied tensile stress
$\sigma_\mathrm{ext}$ amounts to
\begin{equation}
  \tau_\mathrm{SB} = \frac{1}{2} \sigma_\mathrm{ext}.
\end{equation}
This stress acts mainly on the shear-band front, allowing us to write
\begin{equation}
  F_\mathrm{ext} \approx \tau_\mathrm{SB} \times l \times h_\mathrm{SB}
               = \frac{\sigma_\mathrm{ext}}{2} \, l \, h_\mathrm{SB},
\end{equation}
where $h_\mathrm{SB}$ is the width of the shear band.  At the moment
that the shear band hits the precipitate, the force $F_\mathrm{ext}$
must be equal to a reaction force $F_\mathrm{back}$ from the
precipitate (\textit{actio est reactio}). Using the projected
precipitate area $A$ (cf.\ Fig.~\ref{fig:model-geometry}), we
can convert that force into a normal stress:
\begin{equation}
  \sigma_A^n = \frac{F_\mathrm{back}}{A/2}.
\end{equation}%
\begin{figure*}
  \centering
  \includegraphics[]{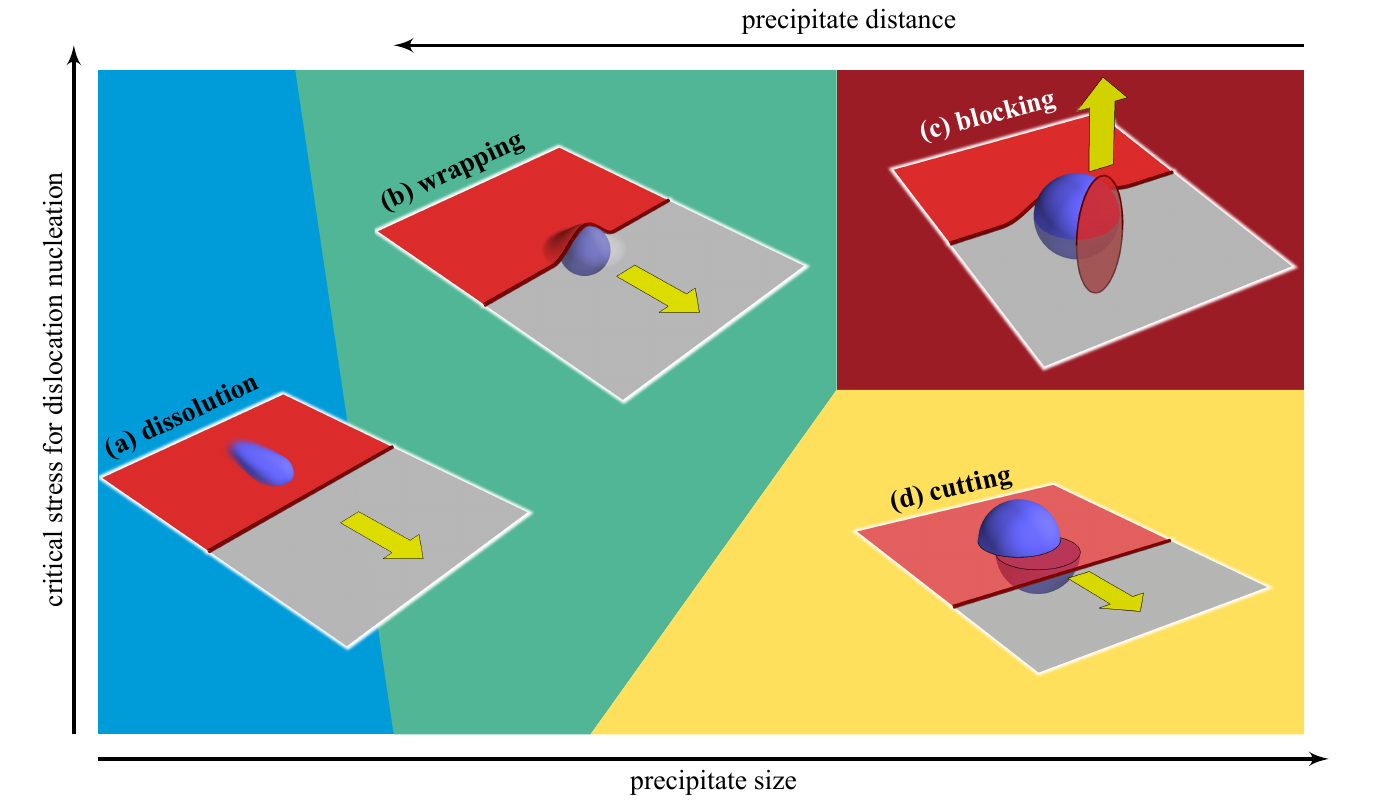}%
  \caption{Schematic view of different mechanisms for the interaction
    of a shear band with a crystalline precipitate.  With increasing
    precipitate sizes, the dissolution of the precipitate is first
    replaced by the wrapping mechanism.  Depending on the critical
    stress for dislocation nucleation, wrapping is replaced by
    blocking or cutting.  Wrapping can be favored by increasing
    the precipitate distance.}
  \label{fig:schema-conclusion}
\end{figure*}%
We assume that $F_\mathrm{ext}$ predominantly acts on one half of the
obstacle (area $A/2$), which is supported by the deformation pattern
of the plastically deformed particle
(Fig.~\ref{fig:plastic-precipitate} and
Video~\ref{fig:plastic-precipitate}).  This back stress results in a
shear stress $\tau_\mathrm{wrap} \approx 0.5\sigma_A^n$ in the plane
of the wrapping shear band (green shear band in
Fig.~\ref{fig:model-geometry}). With $F_\mathrm{back} =
F_\mathrm{ext}$, it is
\begin{equation}
  \sigma_A^n = \frac{F_\mathrm{ext}}{A/2}
             = \frac{\sigma_\mathrm{ext} \, l \, h_\mathrm{SB}}{A}
             = 2\tau_\mathrm{wrap}.
\end{equation}
$\tau_\mathrm{wrap}$ must surpass a critical value
$\tilde{\tau}_\mathrm{wrap}$ to allow the initiation of the wrapping
mechanism; otherwise the shear band simply stops propagating.  This
is not observed in our simulations, suggesting that
$\sigma_\mathrm{ext}$ at yield is greater than
\begin{equation}
  \tilde{\sigma}_\mathrm{ext}
  = \frac{2 A \tilde{\tau}_\mathrm{wrap}}{l \, h_\mathrm{SB}}.
\end{equation}
The competing mechanism, blocking the shear band and nucleating a new
one (red shear band in Fig.~\ref{fig:model-geometry}), can simply be
expressed by a critical shear stress $\tilde{\tau}_\mathrm{nucl}$.
Because of the low temperature in the simulation and a stress close to the
yield stress, we consider only the athermal case and do not invoke a
nucleation term which takes into account the relative volume of the
interface. With $\tau_\mathrm{nucl} = 0.5 \sigma_\mathrm{ext}$, the
transition from wrapping to blocking takes place where
\begin{align}
  \label{eq:transition}
  &2\tilde{\tau}_\mathrm{nucl} = \tilde{\sigma}_\mathrm{ext}
  = \frac{2 A \tilde{\tau}_\mathrm{wrap}}{l \,
    h_\mathrm{SB}}, \quad \text{giving} \\
  \label{eq:relation-lambda}
  &\frac{A}{l} = \Lambda_\mathrm{crit}
   = \frac{\tilde{\tau}_\mathrm{nucl}}{\tilde{\tau}_\mathrm{wrap}} h_\mathrm{SB}.
\end{align}
This derivation also works in the case of externally applied shear
stress, by replacing $\sigma_\mathrm{ext}$ with $2\tau_\mathrm{ext}$.

Assuming that nucleating a new shear band at the interface and
propagating the wrapping shear band along the interface have similar
critical stresses, we can simplify Eq.~\ref{eq:relation-lambda} to
\begin{equation}
  \Lambda_\mathrm{crit} \approx h_\mathrm{SB}.
\end{equation}
Shear bands in Cu-Zr-based glasses have widths of around
$10\,\mathrm{nm}$,\cite{Ritter2011} which fits to the
$\Lambda_\mathrm{crit} = 12.65\,\mathrm{nm}$ observed in our model
systems. As stated earlier, the wrapping mechanism becomes more
favorable again if the shear band is longer before it hits the
precipitate. The reason is that this gives the shear band time to
deviate slightly from its path, so that it does not hit the
precipitate centrally, thus effectively reducing $A$. In practice, this
is not a big problem, as the free shear-band length is constrained to
approximately $l$ anyway due to the distribution of precipitates in the
sample.  While this derivation is only approximate, it seems to be
sufficient to explain the observed phenomena and guide future efforts
in tuning the mechanical behavior of crystal--glass
composites. Furthermore, it is easily possible to explain the fourth
mechanism, a slip transfer into the crystalline phase.  The critical
stress for heterogeneous nucleation of a dislocation in the
precipitate $\tilde{\tau}_\mathrm{disl}$ must be provided by the shear
band via $\tau_\mathrm{SB} = 0.5 \sigma_\mathrm{ext}$. If
$\tilde{\tau}_\mathrm{disl} < \tilde{\tau}_\mathrm{nucl}$, we can
simply replace $\tilde{\tau}_\mathrm{nucl}$ by
$\tilde{\tau}_\mathrm{disl}$ in Eqs.~\ref{eq:transition} and
\ref{eq:relation-lambda}, thereby replacing the blocking mechanism
with the plastic deformation of the precipitate. A lowered
$\tilde{\tau}_\mathrm{disl}$ also lowers $\Lambda_\mathrm{crit}$.

A simple deflection of the shear band is not observed and seems
unlikely, as any deviation of the shear band from its path leads to a
reduced resolved shear stress and thereby to a driving force to put it
back ``on track.''  Contrary to the experiment, a change of shear-band
path towards the precipitate was also not observed.  Because the
precipitates in our simulations are inserted artificially and do not have
a large stress field around them, this seems reasonable. In thermally
grown precipitates, a stress field due to density mismatch between
glass and crystal seems likely. Still, due to the geometry of the MD
simulations, the shear band has two equivalent propagation pathways
from the notch but always chooses the one leading towards the
precipitate.  This seems to be a weaker form of the attraction observed
experimentally.

With these results, we can attempt an explanation of the
experimentally observed phenomena. First of all, the blocking of the
shear band is a one-to-one correspondence between simulation and
experiment. Comparing Fig.~\ref{fig:tem-two-precs-with-sb}(e) with
Fig.~\ref{fig:blocking}, we can see that the path of the shear band
looks identical. The shear band wraps partly around the precipitate
but is then blocked and does not propagate. Contrary to the
simulation, no fully formed shear band but only a small
shear-band-like region appears at the opposite crystal--glass
interface. This may be a result either of the more complex stress
state in the experiment or the fact that other precipitates are
available at which the new shear band may nucleate. For the winding
shear-band path, we can now exclude a simple deflection as
discussed above. A possible explanation would be the concurrent
nucleation of nascent shear bands at the crystal--glass interfaces
which grow together into a single mature shear band. While the
interfaces are known to be nucleation sources for shear
bands,\cite{Albe2013,Zaheri2014,Wang2014b} our simulations show that
the nucleation of a shear band at a stress concentrator like a notch
or a crack always takes precedence to nucleation at interfaces or
surfaces. The shear bands shown in the TEM images all originate from
crack tips, making it unlikely that the shear band shown in
Fig.~\ref{fig:tem-sb-pattern} consists of several concurrently
nucleated shear bands. This leaves the explanation that this winding
path is a series of subsequent blocking and renucleation events.

The observed $\Lambda_\mathrm{crit}$ corresponds to precipitate
diameters somewhere between $20\,\mathrm{nm}$ and $40\,\mathrm{nm}$,
depending on the interparticle distance.  This critical diameter is
on the order of magnitude reported in several experimental studies for
twinning in B2 crystals in Cu-Zr-Al-based metallic glasses of
$20\pm5\,\mathrm{nm}$.\cite{Pauly2010a,Kuo2014} It also fits an
experimental work on Al-based glasses, where crystallites growing
during deformation are sheared apart when they reach a critical size
of about $10\,\mathrm{nm}$.\cite{Hebert2006} Cu$_{50}$Zr$_{45}$Ti$_5$
metallic glasses exhibit a critical size of about $9\,\mathrm{nm}$ for
twinning of B2 precipitates.\cite{Wang2014a} The interparticle
distances in these experiments are on the same order of magnitude as
for our simulations.  This supports an explanation of a transition
from wrapping to slip transfer in these systems.

Figure~\ref{fig:schema-conclusion} summarizes the competition between
the different mechanisms.  Mechanical dissolution of the crystalline
particles occurs only if their size is comparable to the shear-band
size.\cite{Lund2007} With further increasing precipitate sizes, the
shear band can still wrap around the obstacle until the size reaches a
threshold value. This critical size also depends on the precipitate
distance, as discussed before, expressed in the parameter
$\Lambda_\mathrm{crit}(d,l)$. If wrapping is no longer possible, a
precipitate which reacts only elastically to the applied stress will
block the shear band. If the precipitate is susceptible to plastic
deformation, slip transfer into the precipitate will take place.

Concerning the mechanical performance of such \textit{in situ}
composites with crystalline precipitates that originate from
nucleation and growth within the glass, the current results suggest
that the discussed geometrical effects serve to improve the
macroscopic mechanical performance.  None of the presented mechanisms
seem to inhibit the percolation of a critical shear band, yet,
catastrophic slip along a shear band leading to complete failure is
delayed in the case of winding shear bands due to the increase of the
shear-band path length as well as the raised activation barriers for
slip along shear bands that have a more complex topology.  The
wrapping mechanism does not pose an obstacle to shear-band
propagation, but can be avoided by appropriate adjustment of the
crystalline volume fraction and precipitate diameter. ``Soft''
precipitates additionally open possibilities to adjust the plastic
deformation by participating in it.  Consistently, by increasing the
volume fraction of the ductile crystalline phase, the constraints on
shear-band propagation can be increased, immediate failure can be
prevented (cf.\ Ref.~\onlinecite{Hofmann2008}), and the composite
displays macroscopic mechanical behavior according to a mixing rule,
further allowing one to tailor the properties.

\section{Conclusions}
\label{sec:conclusions}

Using TEM imaging, we observe shear-band bending around or close to
precipitates, an attraction of shear bands to the precipitates, and
shear bands being blocked by precipitates. MD simulations reveal that
the shear-band bending is most likely the result of the subsequent
blocking and renucleation of shear bands. Moreover, we identify shear
bands wrapping around precipitates and slip transfer into the
crystalline phase. By describing the competition between the critical
stress for wrapping, the nucleation of a new shear band, and the
nucleation of dislocations in the crystal, we could derive a mechanism
map for metallic glasses with nanocrystalline precipitates. This
detailed description of shear-band propagation not only helps to
understand the mechanical failure of these composites but also aids in
tuning them.

\begin{figure}[b]
  \centering
  \includegraphics{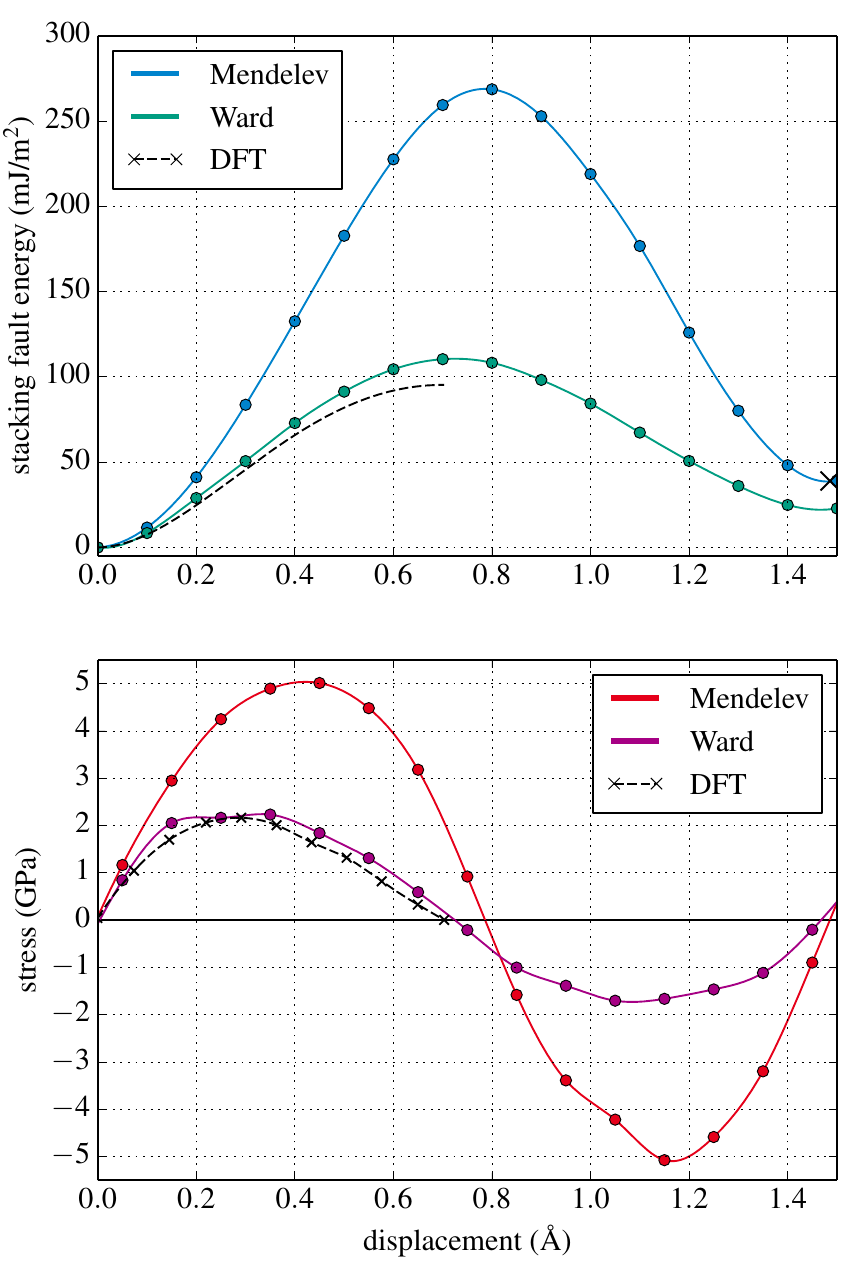}
  \caption{Generalized stacking-fault energies (top) and the resulting
    shear stresses (bottom) for fcc copper.  The DFT values for the
    stress curve are from Ogata \textit{et
      al.},{\protect\ignorecitefornumbering{\cite{Ogata2002}}} and the
    corresponding stacking-fault energies are approximated by using a
    numerical integration of the stress data.}
  \label{fig:GSF}
\end{figure}

\begin{acknowledgments}
  The authors thank J.~B\"unz and M.~Gerlitz for help with sample
  preparation and DSC characterization.
  Financial support by the Deut\-sche For\-schungs\-ge\-mein\-schaft
  (DFG) through project grants nos.\ AL 578/13-1, AL 578/6-2, and WI
  1899/12-1 is gratefully acknowledged.
  Computing time was granted by the John von Neumann Institute for
  Computing (NIC) and provided on the supercomputer JUROPA at J\"ulich
  Supercomputing Centre (JSC), as well as by the Gauss Centre for
  Supercomputing (GCS) through the NIC on the GCS share of the
  supercomputer JUQUEEN at JSC. GCS is the alliance of the three
  national supercomputing centers HLRS (Universit\"at Stuttgart), JSC
  (Forschungszentrum J\"ulich), and LRZ (Bayerische Akademie der
  Wissenschaften), funded by the German Federal Ministry of Education
  and Research and the German State Ministries for Research of
  Baden-W\"urttemberg, Bayern, and
  Nord\-rhein-West\-fa\-len. Additional computing time was made
  available by the Technische Universit\"at Darmstadt on the
  Lichtenberg cluster.
\end{acknowledgments}

\appendix*

\section{Generalized stacking-fault energy in fcc copper}
\label{sec:GSF}

The generalized stacking-fault energy in fcc copper is calculated by
using both Mendelev \cite{Mendelev2009} and Ward
\cite{Zhou2004,Ward2012preprint} potentials. Density-functional theory
(DFT) calculations from Ogata \textit{et al.}  \cite{Ogata2002} using
the same method are used for comparison.  This is plotted in
Fig.~\ref{fig:GSF}.  The Ward potential much more accurately describes
the stacking-fault energy than the Mendelev potential, which has a
critical stress for homogeneous dislocation nucleation which is more
than two times too high.  \vfill

\bibliography{literature.bib}

\end{document}